\documentclass[12pt]{article} 

\newcommand{\manuscript}{Microstructural and texture evolution of Copper-(Chromium, Molybdenum, Tungsten) composites deformed by high-pressure-torsion}
\newcommand{\shortmanuscript}{Hpt of Cu-(Cr/Mo/W)}
\newcommand{\shortauthors}{Rosalie, Ghosh, Guo, Renk, Zhang}
\usepackage[numbers,sort&compress]{natbib}

\usepackage[left=2cm,right=2.5cm,top=2cm,bottom=2cm]{geometry}
\usepackage{fancyhdr}
	\fancypagestyle{plain}{ %
	\fancyhf{} 
	\lhead{\textit{\shortmanuscript}}
	\rhead{\shortauthors}
	\rfoot{\thepage}
	}

\usepackage{amsbsy} 
\usepackage{graphicx}
\usepackage{subfigure}
\usepackage[version=3]{mhchem} 
\usepackage{multirow} 
\usepackage{multicol}
\usepackage{booktabs}	
\usepackage{setspace}
\usepackage{xspace}
\usepackage{comment}
\usepackage{amssymb}

\title{\manuscript}

\author{Julian M. Rosalie$^{a}$
\and Pradipta Ghosh$^{a}$
\and Jinming Guo$^{a}$,
\and Oliver Renk$^{a}$,
\and Zaoli Zhang$^{a}$.
}	
\date{}
\vspace{6pt}

\begin{document}

\maketitle
\noindent
$^a$Austrian Academy of  Sciences, Erich Schmid Institute of Materials Science, A-8700 Leoben, Austria.\\
\begin{abstract} 
Cu-refractory metal composites containing Cr, Mo or W were subjected to severe plastic deformation using room temperature high-pressure torsion (HPT). A lamellar microstructure developed in each of the composites at equivalent strains of $\sim$75. The refractory metals developed $\{hkl\}\langle111\rangle$ fibre textures with a slight tilt to the tangential direction. This texture was stronger and more clearly defined in Mo and W than in Cr.

By applying additional HPT deformation to these samples, perpendicular to the original shear strain, it was found that the lamellar structure of Cu30Mo70 and Cu20W80 (wt.\%) composites could be retained at high equivalent strains and the refractory layer thickness could be reduced to 20--50\,nm in Cu20W80 and 10-20\,nm in Cu30Mo70. Although neighbouring regions of the microstructure were aligned and there was evidence of local texture in both composites, the bulk texture of Cu30Mo70 became weaker during this second step of HPT deformation. This was attributed to the refractory metal lamellae being discontinuous and imperfectly aligned.

This work shows that it is possible to form ultrafine composites of Cu-group VI refractory metals via high-pressure torsion, with namolamellar structures being possible where there is a sufficient  volume fraction of Mo or W.
\end{abstract}


\section{Introduction}

Copper-refractory metal composites are important industrial materials consisting of varying proportions of copper and one or more of the refractory metals, namely Nb, Ta, Cr, Mo, W and Re. These composites are employed in high temperature, high current applications such as heat sinks \cite{AguilarGuzman2011},  electrodes in thermoelectric devices \cite{AguilarGuzman2011,KumarJayasankar2015,BaiLi2015} and in aerospace components \cite{WadsworthNieh1988} which demand high thermal and/or electrical conductivity of Cu and the capability to operate at higher temperatures than copper alone can sustain. The high melting points (from 1907$^\circ$C(Cr)--3422$^\circ$C (W))\cite{Cockeram2002} and low thermal expansion coefficients (4.5--7.3\,$\mu$m\,m$^{-1}$\,K$^{-1}$)\cite{CrcHandbook2002Thermophysical} of the refractory elements provide the required high temperature strength and dimensional stability. 

Nanostructured forms of these composites have been produced by severe plastic deformation (SPD). While techniques such as accumulative cold drawing and bundling (ADB) and equal channel angular extrusion (ECAE) have been used \cite{BuetDubois2010,BuetDubois2012}, the best studied examples are the Cu-Nb composites produced by accumulative roll-bonding (ARB)\cite{Beyerlein2013,CarpenterZheng2013,EkizLach2014,MisraHoagland2004,MisraDemkowicz2007,CarpenterVogel2012} 
by which it has been possible to produce lamellar structures with widths of 10--20\,nm\cite{Beyerlein2013,MisraHoagland2004,MisraDemkowicz2007,CarpenterVogel2012}.

The lamellar microstructure endows these composites with excellent thermal stability \cite{MisraHoagland2004, PrimoracAbad2015} by restricting diffusion through the alternating layers. Fracture toughness studies on lamellar composites have also shown a strong anisotropy due to crack deflection along the lamellae\cite{HohenwarterVoelker2016}. Thus, the ability to develop nanolamellar composites of refractory elements in Cu matrix will open-up the possibility for developing high strength, tough and thermally stable components for high temperature applications. 

The production of such nanolaminates relies on the high ductility of the group V refractory metals, with ductile to brittle transformation temperatures (DBTTs)   of $\sim$77\,K  \cite{AdamsRoberts1960,Johnson1960} and $\sim4$\,K \cite{Buckman2000} for Nb and Ta, respectively. 
Unfortunately the group VI refractory metals (Cr,Mo,W) are more brittle, with DBTTs in the range of 553--573\,K for 99.96\% purity Cr \cite{WadsackPippan2001},  about 373\,K \cite{ShieldsLipetzky2001} for Mo and 400\,K for W\cite{Stephens1972}.

High-pressure torsion (HPT) is a technique for applying high strains under quasi-hydrostatic conditions, thus permitting the deformation of brittle materials. 
This has been applied to Cu-Cr\cite{BachmaierRathmayr2014,GuoRosalie2017,GuoRosalie2017a} 
Cu-W\cite{SabirovPippan2005,SabirovPippan2007,KraemerWurster2014} and Cu-Mo \cite{RosalieGuo2017} composites. 
Extensive deformation of Cu-Cr \cite{GuoRosalie2017} results in the formation of equiaxed, nanometer scale grains. A similar microstructure was reported after two-step deformation of Cu50Mo50\cite{RosalieGuo2017}.  In contrast, a Cu30Mo70 (wt.\%) composite retained a lamellar structure with layer widths of $\sim$5\,nm for Cu and $\sim$10--20\,nm for Mo \cite{RosalieGuo2017}. 
It appears that Cu-10wt.\%W and Cu-25wt.\%W composites also retained lamellar structures for strains up to at least 900\cite{KraemerWurster2014}.  However, each of these studies were conducted using different sample sizes, pressure and applied strain, making it difficult to compare various Cu-refractory metal composites. 

This work sets out to provide a comparison of the microstructural evolution of copper composites containing a group VI refractory element (Cr, Mo,W) during equivalent room-temperature HPT deformation. In particular it examines the extent of microstructural refinement, the formation and decomposition of lamellar microstructures and the development of crystallographic texture during deformation. 

\section{Experimental details}

The materials used in this work were provided by Plansee SE, Austria. Cu-Mo and Cu-W composites took the form of liquid-metal infiltrated (LMI) plates, with compositions of Cu30Mo70, Cu50Mo50 and Cu20W80 (All compositions are given in wt.\%). Cu-Cr was in the form of 4\,mm slices sectioned from directionally-solidified Cu57Cr43 billets. 

Step 1 HPT was carried out at room temperature on a 4000\,kN apparatus rotating at 0.0625 revolutions per minute (rpm) under an applied pressure of 4.5\,GPa. The sample thickness, $t$ and number of revolutions, $N$, was adjusted to achieve a nominal von Mises' equivalent strain, $\epsilon_{eq}$ of $\sim$75 at a given radius, $r$, based on the equation~\cite{SaundersNutting1984}:
\begin{equation}
	\epsilon_{eq} \cong \frac{\gamma}{\sqrt{3}} =   \frac{ 2 \pi r N }{ \sqrt{3} t} 
\label{eqn-von-mises}
\end{equation}
Sample were constrained within tool steel anvils of diameter 30\,mm and anvil gap 3.5\,mm (Cu-Mo, Cu-W) or 1.5\,mm (Cu-Cr), corresponding to a nominal sample thickness of 7.5\,mm and 3.5\,mm, respectively. Table~\ref{tab-deformation} lists the deformation conditions  for each composite, and gives the sampling radius and equivalent strain used for XRD analysis.

\begin{table*}[htbp]
	\renewcommand{\thefootnote}{\fnsymbol{footnote}}
	\begin{center}
	\caption{Sample compositions and deformation conditions for step 1 HPT. \label{tab-deformation} The von Mises' equivalent strain $\epsilon_1$ is given for $r'$: the radial distance at which the XRD samples were extracted. (See Figure~\ref{fig-xrd-geometry}.)}
	\begin{tabular}{lp{5ex}p{9ex}p{11ex}p{7ex}p{6ex}}
\toprule
 		&\%\,Cu  &Thick\-ness	&	Rota\-tions	&Radial dist\-ance & Strain	\\
	& 		 			&		$~t$ &	$N$	 & $r'$	& $\epsilon_1$    			\\  
	&	(wt.)		&	(mm)			& 	&	(mm)	&		 \\
\midrule
Cu-Cr	& 57				&3.5	& 5			& 15		&	78									\\
Cu-Mo	& 50				&7.5	& 10			& 15		&	73								\\
		& 30				&7.5	& 10			& 15		&	73									\\
Cu-W		& 20			&7.5	& 20     		& 8		&	77											\\ 	 
\bottomrule
	\end{tabular}

	\end{center}
\end{table*}

\begin{table*}[htbp]
	\renewcommand{\thefootnote}{\fnsymbol{footnote}}
	\begin{center}
	\caption{Sample compositions and deformation conditions for step 2 HPT.  In the case of Cu57Cr43 a single step deformation with a higher number of rotations was used instead, as described in the text. The von Mises' equivalent strains ($\epsilon_1$ and $\epsilon_2$) for step 1 and step 2 deformation are listed for the conditions examined via TEM and synchrotron XRD.
\label{tab-deformation2}}
	\begin{tabular}{lp{5ex}p{9ex}p{11ex}p{7ex}p{6ex}}
\toprule
&\%\,Cu  &Thick\-ness	&	Rota\-tions	&Radial dist\-ance & Strain	\\
	& 		 			&		$t_1:t_2$ &	$N_1:N_2$	 & $r'$	& $\epsilon_{1}:\epsilon_{2}$    			\\  
	&	(wt.)		&	(mm)			& 	&	(mm)	&		 \\
\midrule

Cu-Cr	& 57				& $-$:0.7	&$-$:100					& 	 $-$:3	&	$-$:1,500									\\

Cu-Mo	& 50 &	7.5:0.7	& 10:20\footnotemark[1]			& 8:3		&	50:400								\\
		& 50 &  7.5:0.7	& 10:50\footnotemark[2]			& 	8:3	&	50:1,000								\\
		& 30 &	7.5:0.7	& 10:50			& 	8:3	&	50:1,000									\\
Cu-W	& 20 &	7.5:0.7	& 20:35     		& 8:3		&	100:700											\\ 	 
\bottomrule
\multicolumn{4}{l}{\footnotemark[1]	\footnotesize{Synchrotron XRD}}\\
\multicolumn{4}{l}{\footnotemark[2]	\footnotesize{TEM}}\\
	\end{tabular}
	\end{center}

\end{table*}

Vickers microindentation hardness testing was conducted on cross-sectioned samples of each HPT-deformed composites, using 300\,g or 500\,g loads.
 
The as-received and post-deformation microstructures of each composite were examined using a Gemini 1525 scanning electron microscope (SEM). 
Sample preparation was via mechanical polishing and grinding, with final polishing using a Buehler vibromet polisher. 

\begin{figure}
	\begin{center}
	\includegraphics[width=8cm]{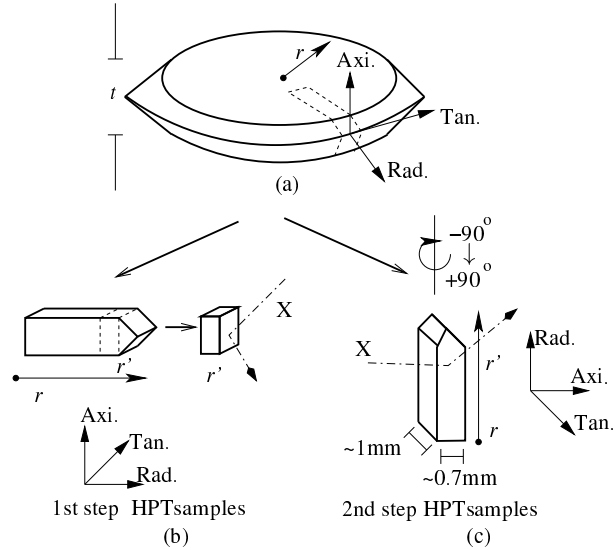}
	\caption{(a) Sample geometry and (b,c) testing geometry for XRD texture samples. (b) Reflection geometry used for first step HPT samples. (b) Transmission geometry used for second step HPT samples.\label{fig-xrd-geometry}}
	\end{center}
\end{figure}

Composites were then subjected to higher-strain HPT deformation to further refine the microstructure. For Mo and W-based composites, the experiments used the two-step method developed by Bachmaier \textit{et. al.} \cite{BachmaierKerber2012}. Cylinders were cut from the deformed 30\,mm diameter Cu-Mo and Cu-W  samples, and sliced into discs with a nominal height of 1\,mm. (The central axis of the the cylinder was at a distance of approximately 8\,mm from the centre of the 30\,mm diameter disc.) These were deformed at a pressure of 7.3\,GPa in anvils with a cavity height of 0.20 or 0.25\,mm. A total of 20--50 rotations of HPT deformation were applied, at a rotation rate of 0.4\,rpm. 
Because the shear plane in step 2 is perpendicular to that in step 1 this method has been shown to be particularly effective for microstructural refinement\cite{BachmaierKerber2012}.
Unfortunately, the 4\,mm thickness of the Cu-Cr starting material did not permit this procedure. Therefore, 8\,mm diameter discs of the as-received material with a nominal height of 1\,mm were deformed to 100 rotations (This was considered acceptable since previous studies on this material showed minimal change in hardness and grain size between deformation of 100 and to 1,000 rotations\cite{GuoRosalie2017a}). Hardness testing was conducted to determine appropriate strain levels for further analysis. The sample size and deformation conditions for step 2 HPT samples are set out in Table~\ref{tab-deformation2}.

Suitably-deformed HPT discs were mechanically ground and polished, dimpled and thinned to perforation using a Gatan 691 precision ion polisher. The electron transparent region corresponded with a radial distance of $\sim$3\,mm. The corresponding strains ($\epsilon_1$:$\epsilon_2$) were $\sim$50:1,000 for Cu-Mo ,  $\sim$100:700 for Cu-W and $\sim$1,500 for Cu-Cr($\epsilon_2$ only) composites. These foils were examined using a JEOL 2100F transmission electron microscope, operating at 200\,kV using high-angle annular dark field scanning TEM (HAADF-STEM). Inner and outer collection angles of 65.51\,mrad and 174.9\,mrad, respectively, were used to provide atomic contrast imaging.

The bulk crystallographic texture of samples deformed till $\epsilon_{eq}\sim75$ (i.e, first step of HPT) were measured using a Rigaku Smartlab system with Cu K$\alpha$ radiation. An approximate $\sim$4\,mm$\times$4\,mm samples normal to the radial direction were used for these measurements. The defocusing corrections due to sample tilts were made using powders or powder compacts of the same size and composition. The texture measurements on the second step HPT deformed samples were performed with synchrotron radiation at the high energy materials science beamline P07 (operated by the Helmholtz-Zentrum Geesthacht) of the PETRA III synchrotron facility at DESY in Hamburg with photon energies of 101\,keV. For this experiment, radial slices with a nominal width of 1\,mm were cut from the HPT discs (see Figure~\ref{fig-xrd-geometry}(c)). The transmitted diffraction rings were measured with a beam size of 0.8\,mm (high) x 1\,mm (wide). The samples were rotated about their radial axis by 180$^\circ$ with an interval of 5$^\circ$ in order to obtain three-dimensional texture information. Further, the intensity of the diffracted beams were corrected for the attenuation. A LaB$_6$ NIST standard was used for calibrating the detector distance and orientation with respect to the sample. For both techniques, the measured pole figures were analysed with Labosoft software. For W and Mo (110), (200) and (211) pole figures were measured while for Cr (200), (211) and (022) pole figures were measured, (due to overlap of the Cu (111) and Cr (110) peaks). The measured pole figures were used to calculate the orientation distribution function (ODF), from which the (110) and (222) projections were recalculated.

\section{Results}

\subsection{Hardness}

\begin{figure}
	\begin{center}
		\subfigure[\label{fig-hardness-1}]{\includegraphics[width=6cm]{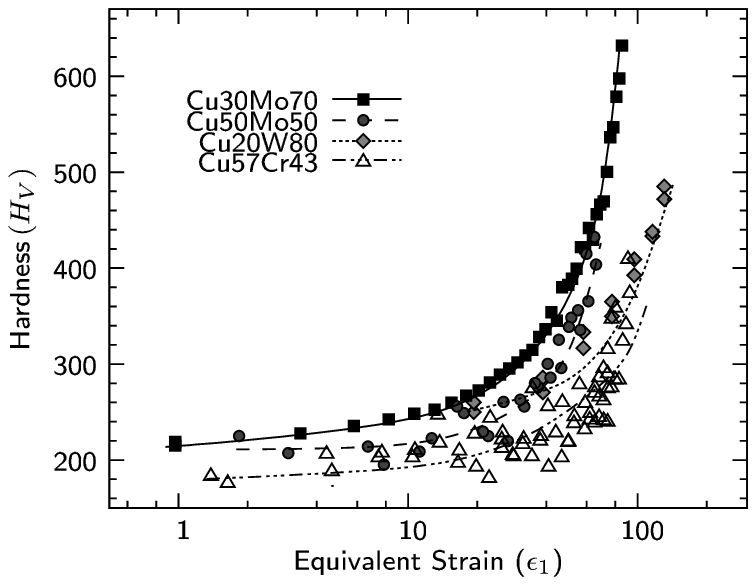}}

		\subfigure[\label{fig-hardness-2}]{\includegraphics[width=6cm]{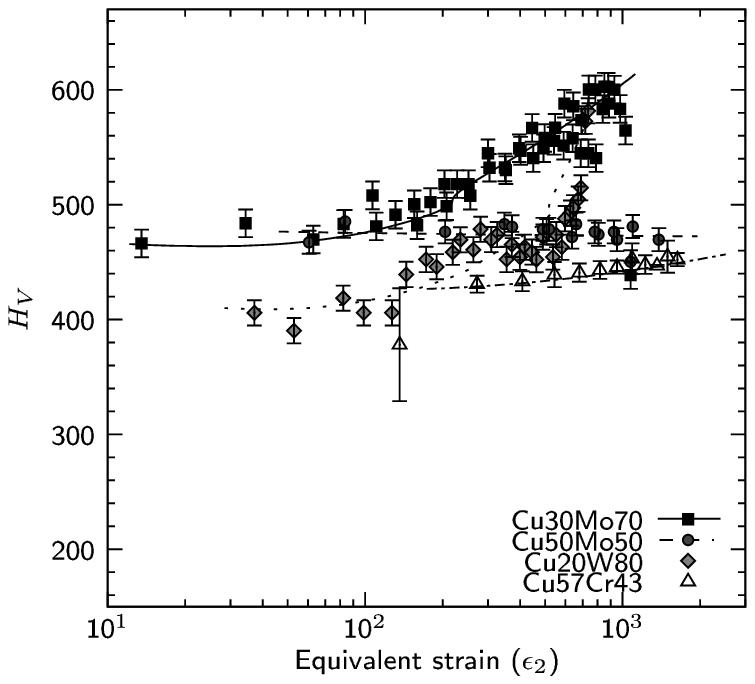}}
		\caption{Hardness data for the HPT-deformed Cu-refractory composites plotted against equivalent strain, $\epsilon$  for (a) first and (b) second step HPT deformation. Lines are shown solely as a guide for the eye. \label{fig-hardness}} 
	\end{center}
\end{figure}

The hardness of the composites after HPT deformation is plotted against  the applied equivalent strain ($\epsilon_{eq}$) in Figure~\ref{fig-hardness}. The change in hardness during the first step of HPT deformation is shown in Figure~\ref{fig-hardness-1}. Each of the composites underwent a rapid increase in hardness commencing at strains of $\epsilon_1\sim$20, with the greatest increases in hardness occurring in the Cu-Mo composites. The SEM and XRD investigations were performed  at a strain value of $\epsilon_1\sim$75. Figure~\ref{fig-hardness-2} shows the hardness of the composites after step 2 deformation. 
The hardness of Cu57Cr43 in Fig.~\ref{fig-hardness-2}  increases gradually and approximately linearly with $\log(\epsilon)$ while that of the Cu50Mo50 composite reaches a saturation in hardness. The hardness of the Cu30Mo70 and Cu20W80 composites continues to increase substantially, indicating that these composites have not reached a saturated state. Although the Cu20W80 composite was harder than Cu30Mo70 in the as-received condition, the latter composite became substantially harder after
 HPT deformation, owing to more extensive microstructural refinement, as is shown in the following section. 

\subsection{As-received condition}
\subsubsection{Initial microstructure}

\begin{figure*}
	\begin{center}
		\hfill
		\subfigure[Cu57Cr43\label{fig-sem-cu57c43r-ar}]{\includegraphics[width=4cm]{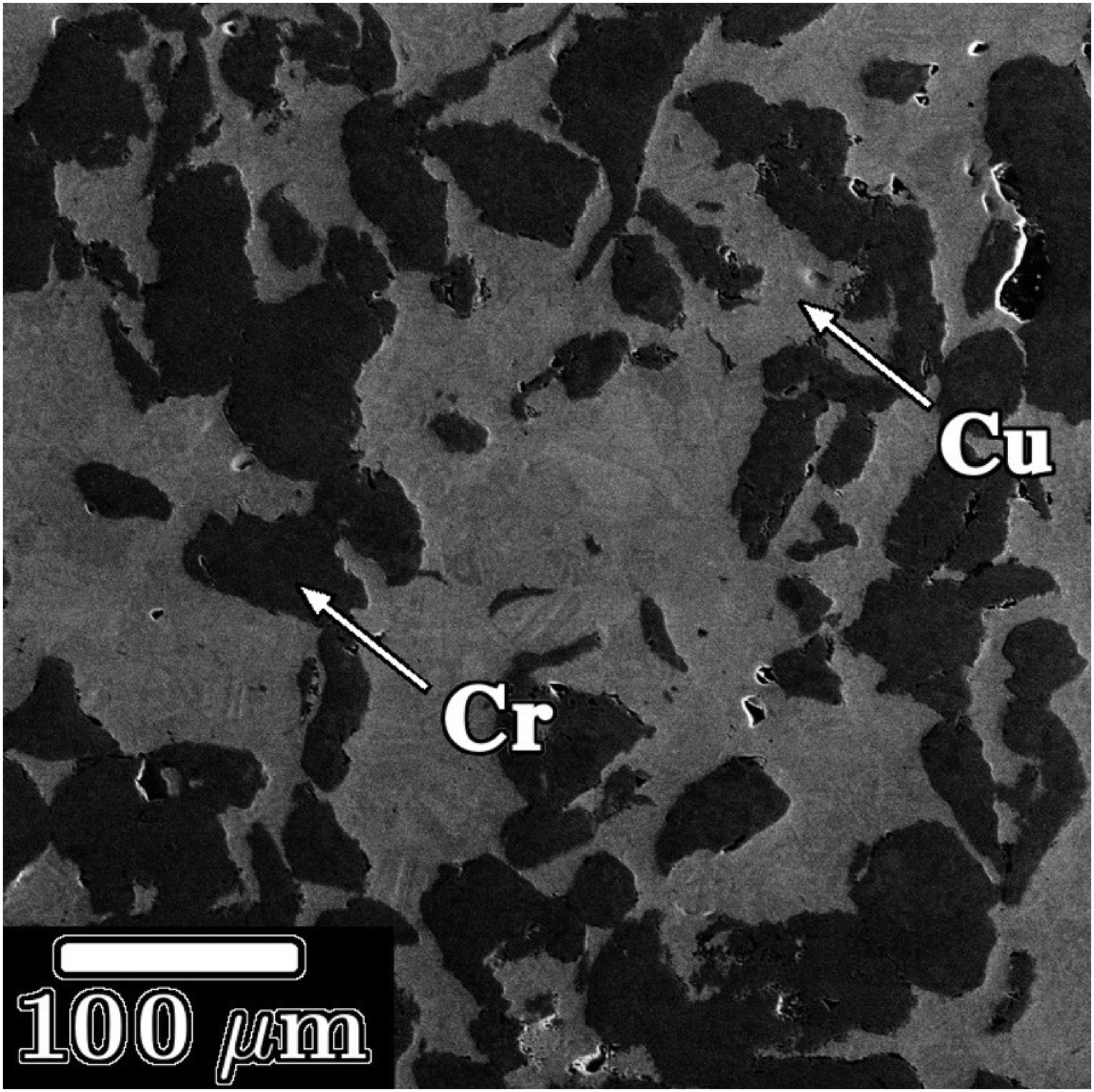}} 
		\hfill 
		\subfigure[Cu50Mo50\label{fig-sem-cu50mo50-ar}]{\includegraphics[width=4cm]{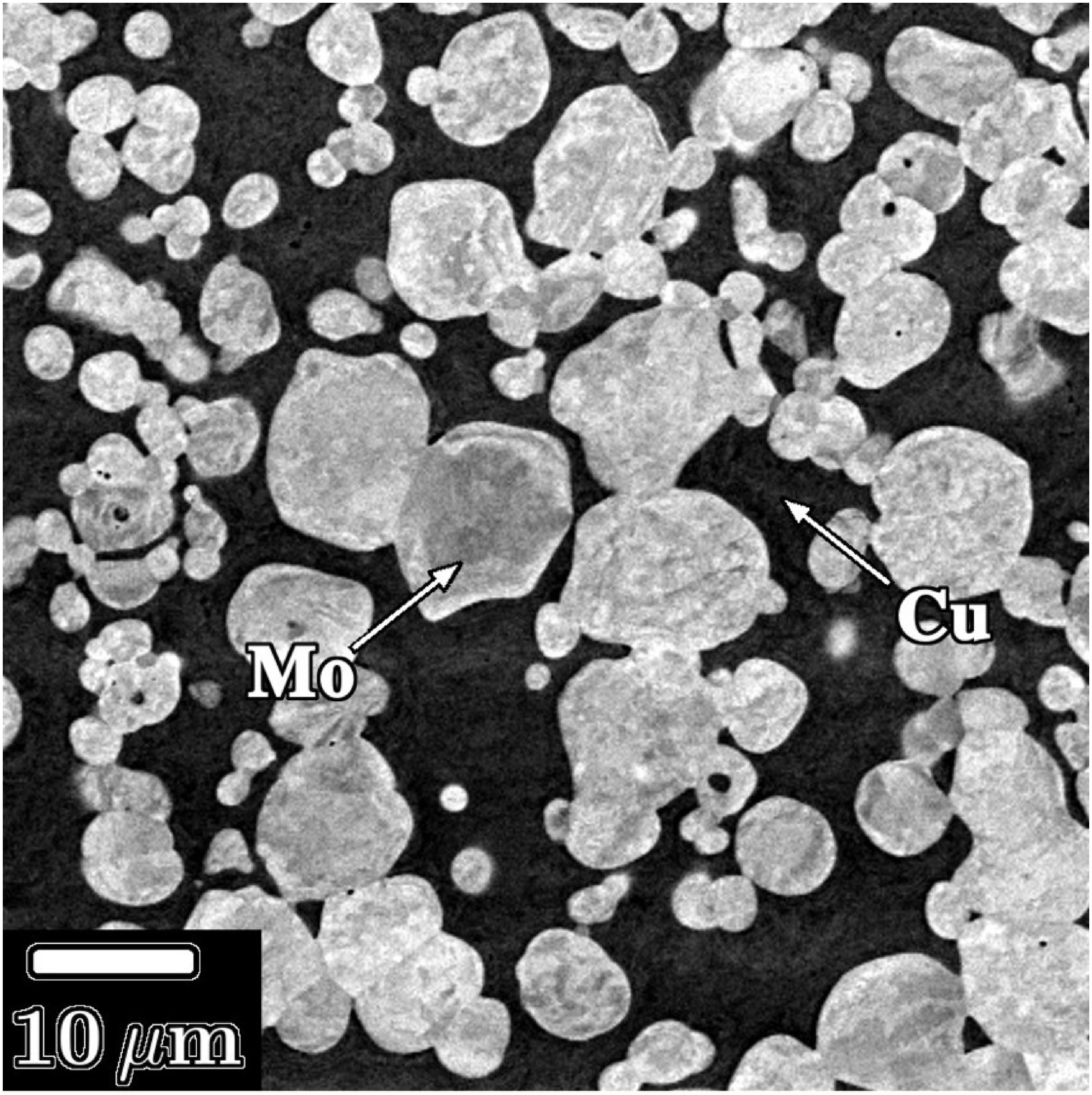}}
		\hfill
		\subfigure[Cu30Mo70\label{fig-sem-cu30mo70-ar}]{\includegraphics[width=4cm]{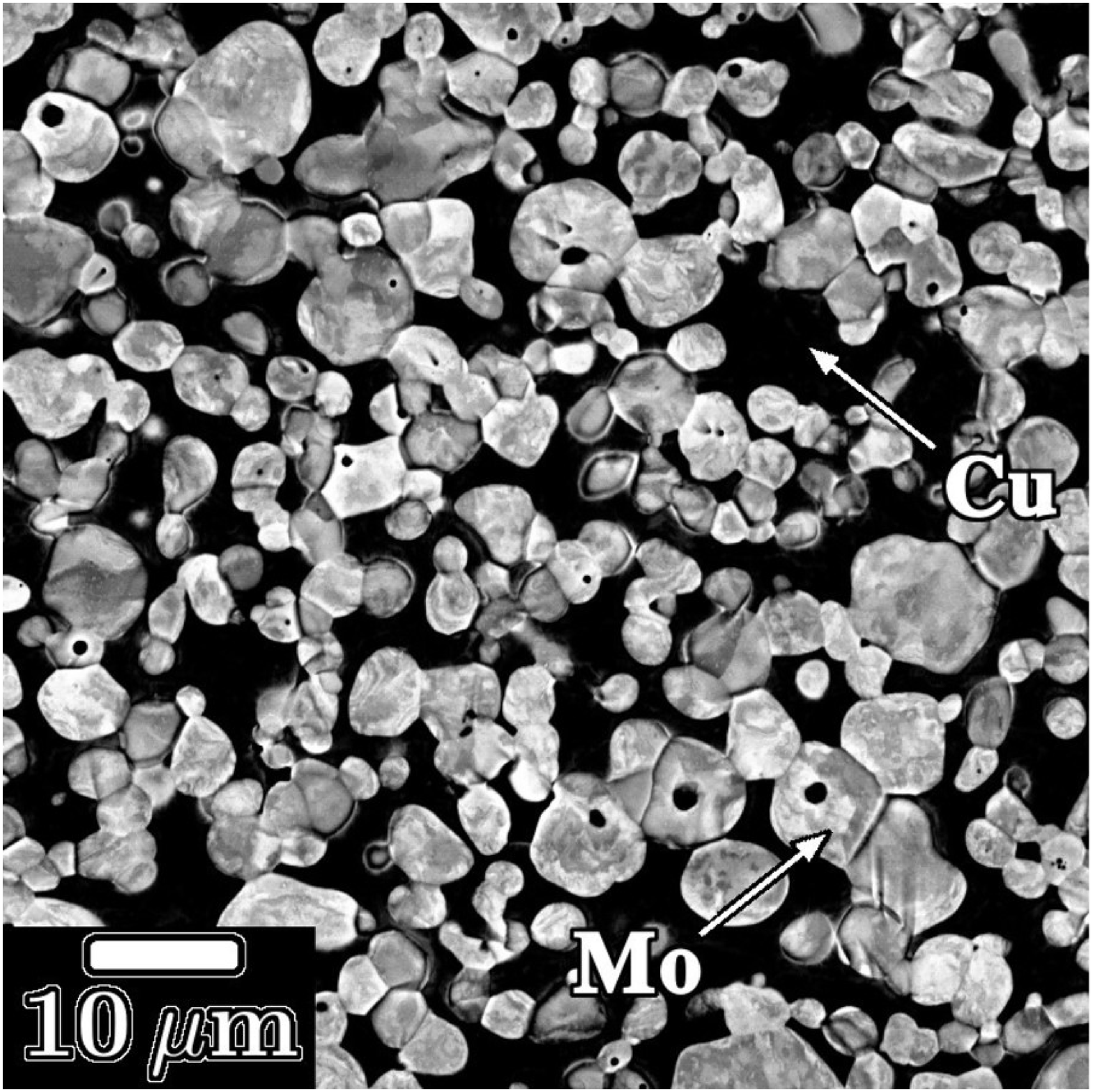}}
		\hfill
		\subfigure[Cu20W80\label{fig-sem-cu20w80-ar}]{\includegraphics[width=4cm]{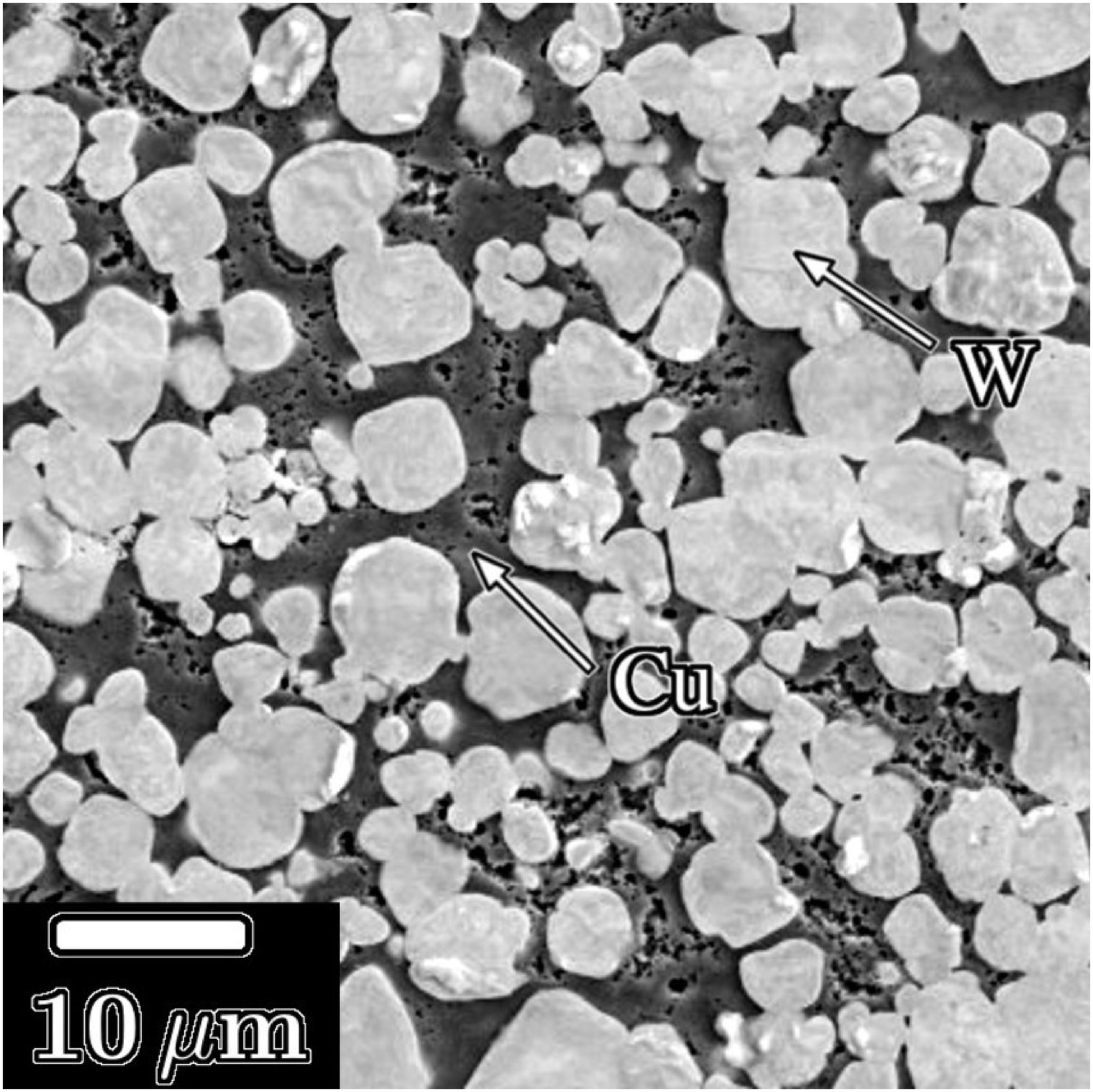}}
		\hfill\ 

		\caption{Backscattered electron micrographs of the as-received Cu-refractory composites. Bright regions indicate the refractory phase, except for Cu-Cr, in which case dark regions indicate chromium.\label{fig-sem-cu-all-ar}}
	\end{center}
\end{figure*}

Backscattered electron micrographs of the as-received composites are provided in Figure~\ref{fig-sem-cu-all-ar}. The Cu-Cr composite (Fig.~\ref{fig-sem-cu57c43r-ar}) had a coarse structure consisting of irregular Cr particles within a Cu matrix. The mean projected area of the Cr particles was 1340\,$\mu$m$^2$ (equivalent in projected area to a spheroidal particle of diameter 41\,$\mu$m) although particles with Feret diameters from 5--100\,$\mu$m were observed. Micron-scale pores are evident, primarily at the interfaces between the two components. The Cu-Mo (Figure~\ref{fig-sem-cu30mo70-ar},\ref{fig-sem-cu50mo50-ar}) and Cu-W (Figure~\ref{fig-sem-cu20w80-ar}) composites contained substantially finer, equiaxed particles of the refractory phase. The average diameter of the refractory metal particles was 3.8\,$\mu$m for Cu30Mo70, 5.0\,$\mu$m for Cu50Mo50 and 3.6\,$\mu$m for Cu20W80.

\subsection{HPT deformed ($\epsilon_1\sim75$)}

\subsubsection{Microstructure\label{sssec-microstructure1}}

\begin{figure}
	\begin{center}
	\hfill\hfill \ce{Near-centre} \hfill 		\includegraphics{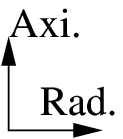} \hfill \ce{Edge} \hfill\hfill \ 

	\hfill
	\rotatebox{90}{\parbox{3.7cm}{\centering Cu57Cr43}}\ 
		\subfigure[]{\includegraphics[width=3.7cm]{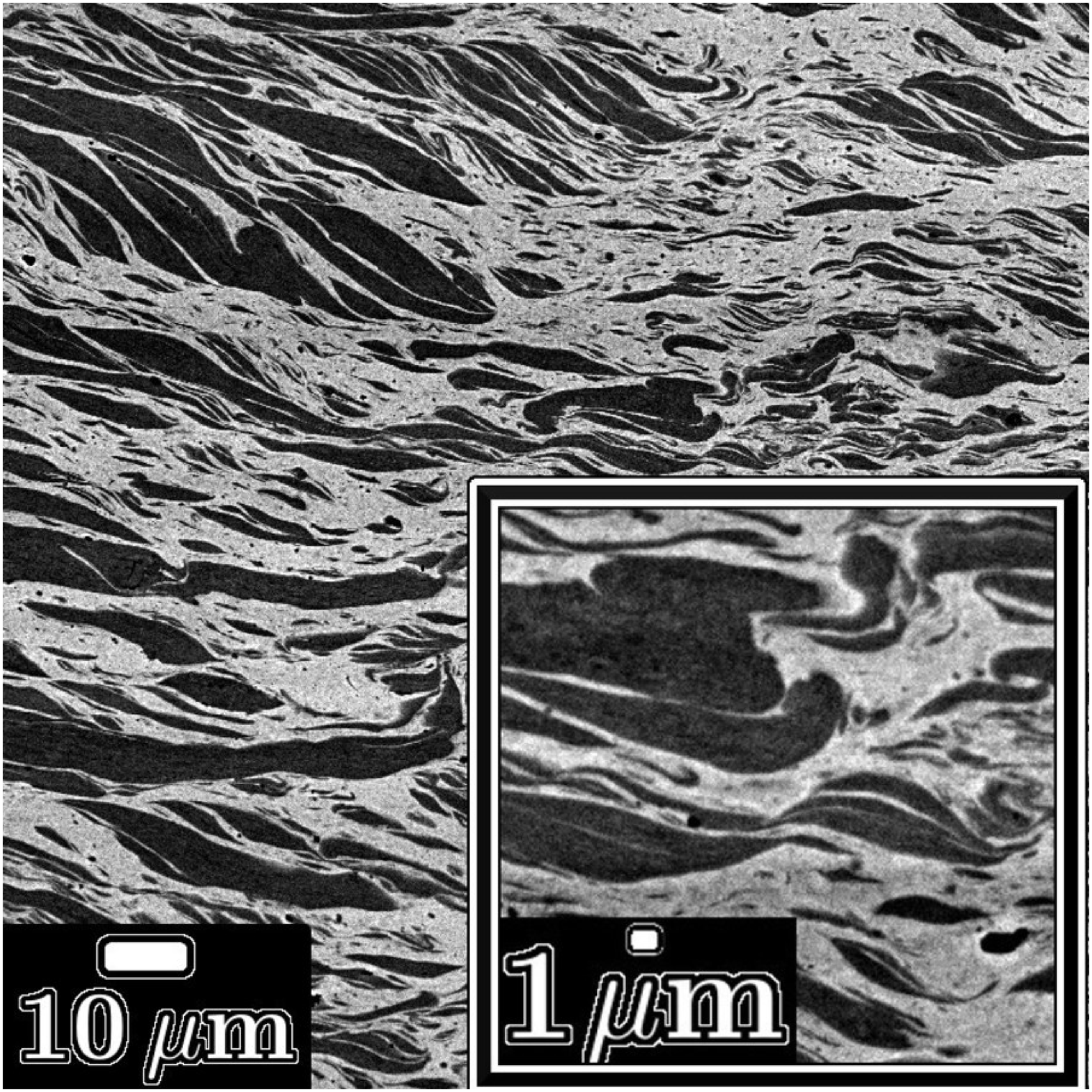}} \hfill 
		\subfigure[]{\includegraphics[width=3.7cm]{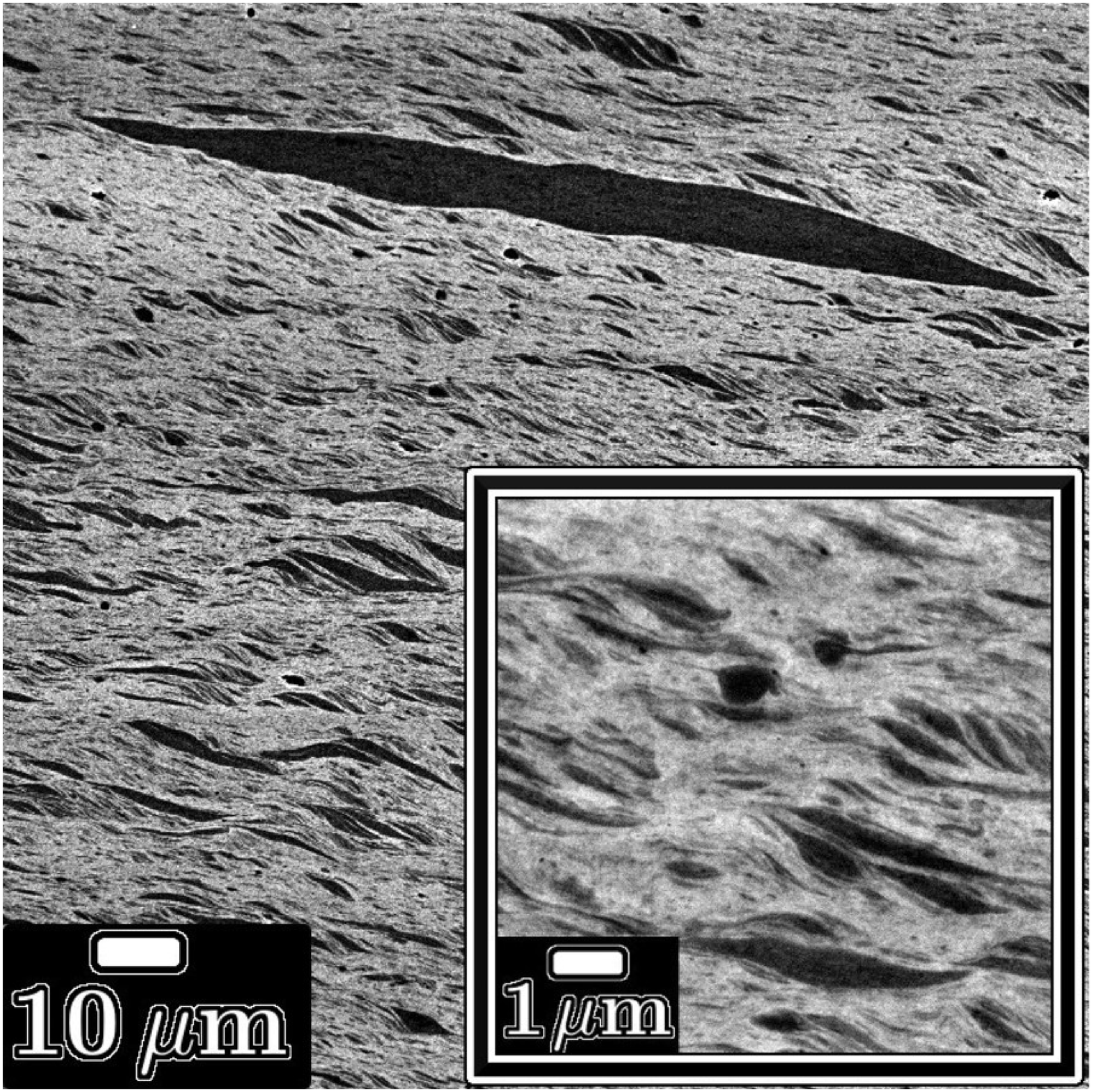}} \hfill\ 

		\hfill
		\rotatebox{90}{\parbox{3.7cm}{\centering Cu50Mo50}}\  
		\subfigure[]{\includegraphics[width=3.7cm]{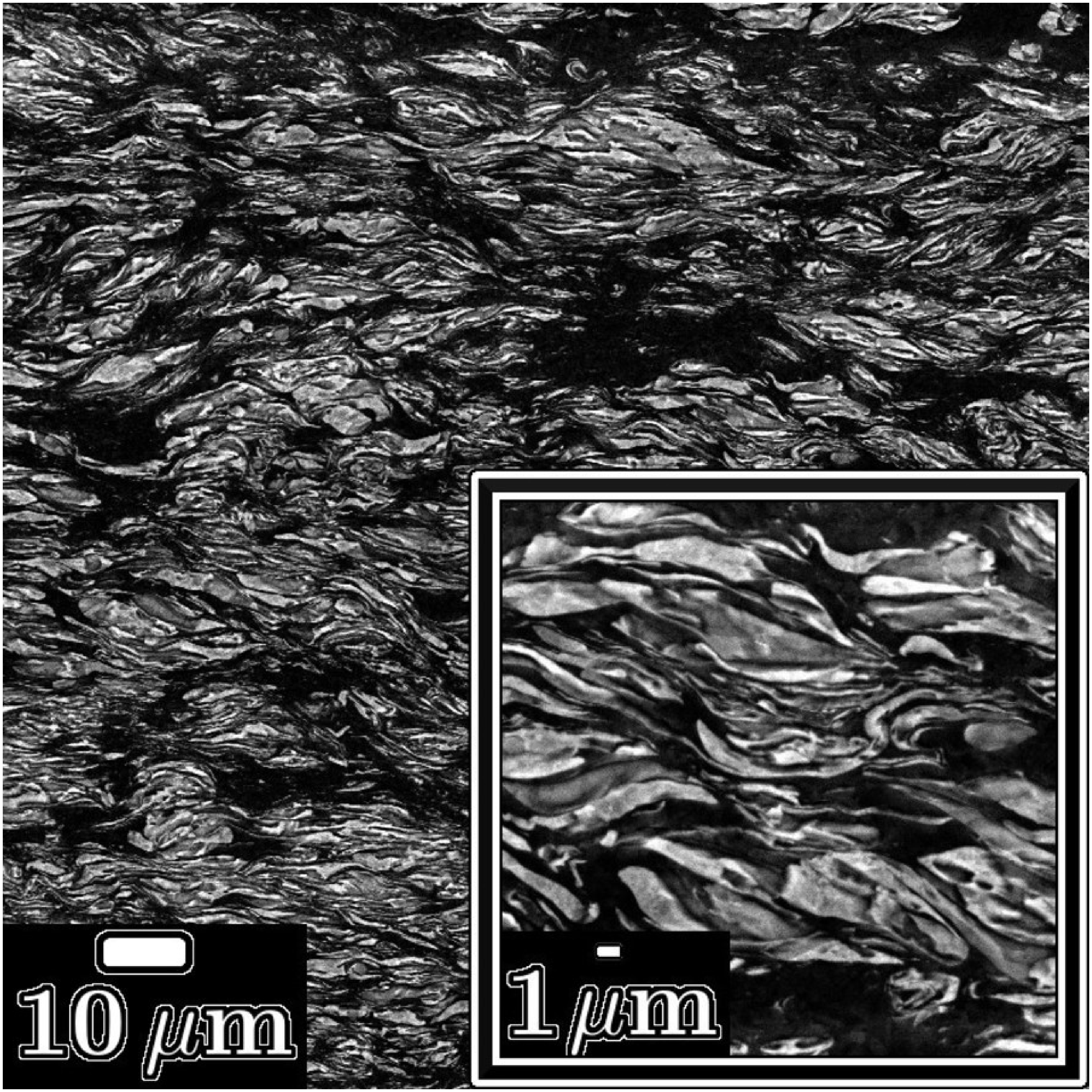}} \hfill 
		\subfigure[]{\includegraphics[width=3.7cm]{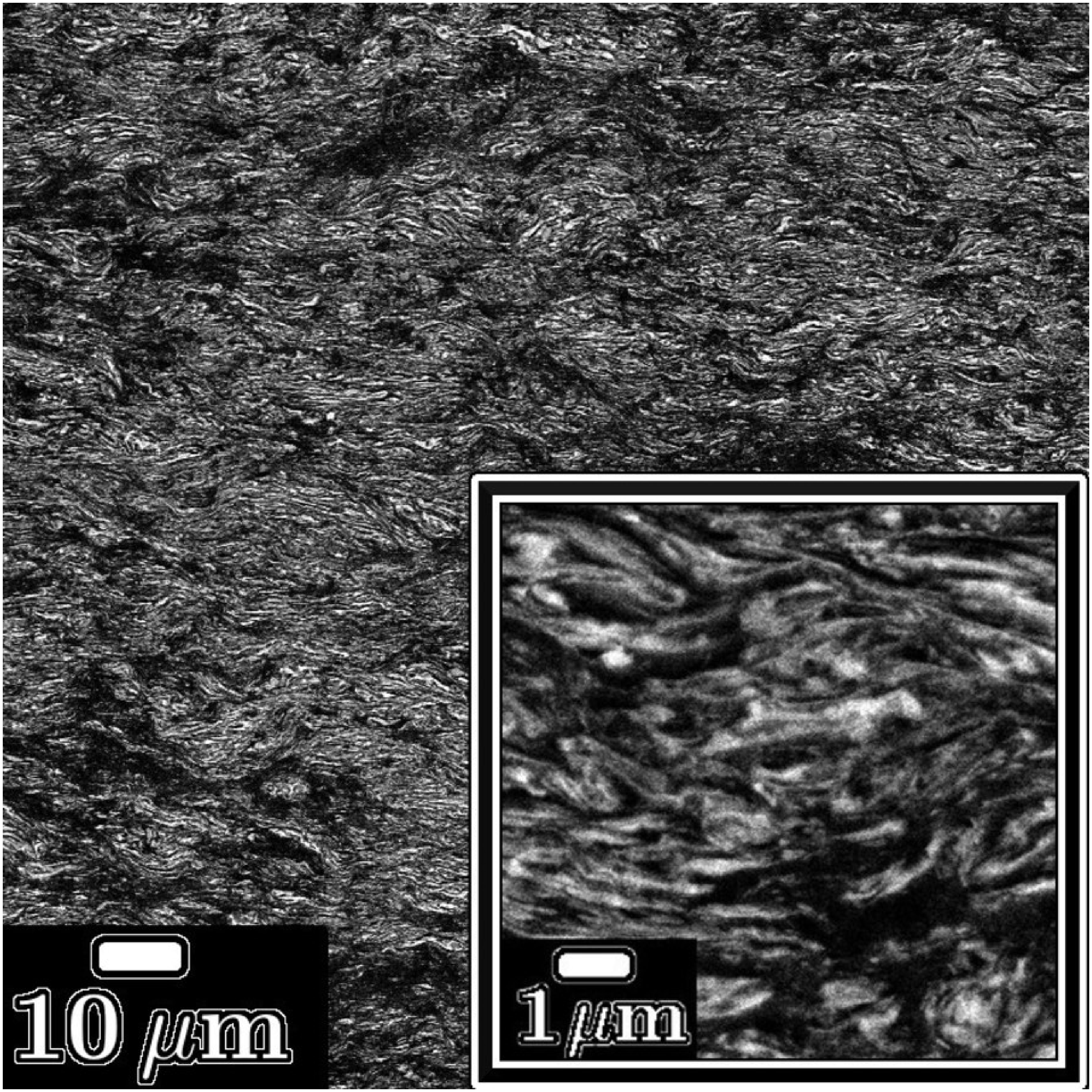}} \hfill\ 

		\hfill
		\rotatebox{90}{\parbox{3.7cm}{\centering Cu30Mo70}}\  
		\subfigure[]{\includegraphics[width=3.7cm]{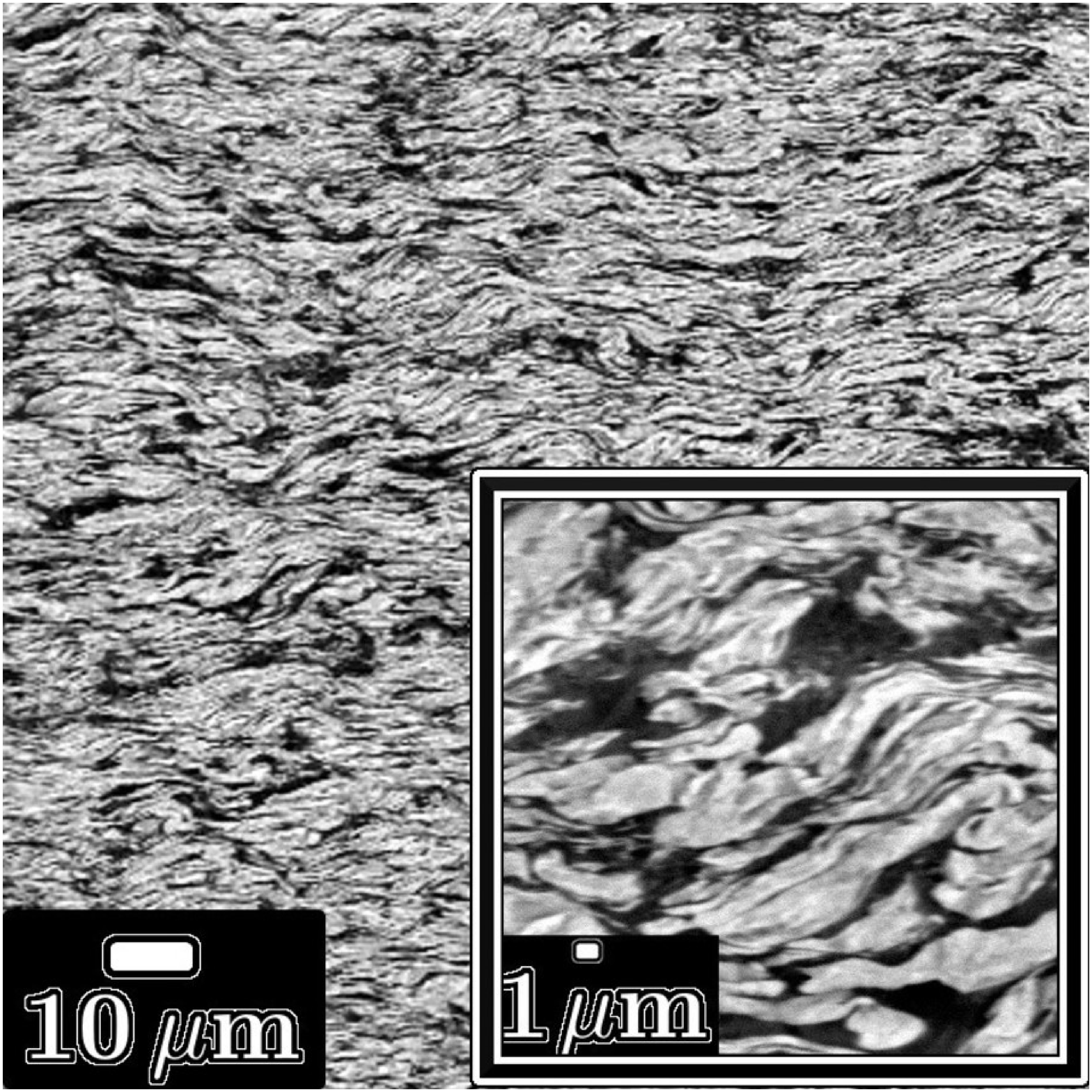}} \hfill 
		\subfigure[]{\includegraphics[width=3.7cm]{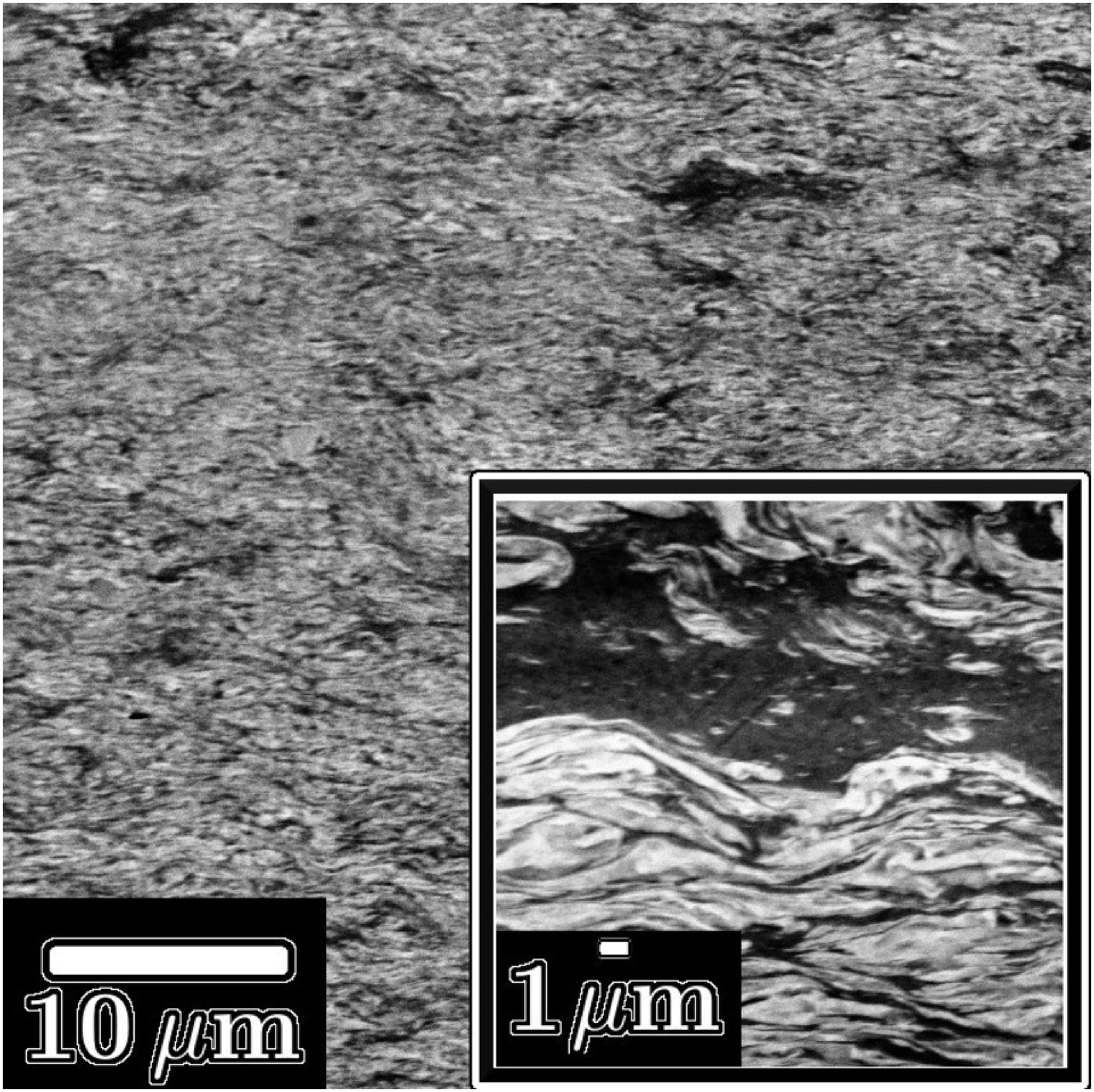}} \hfill\ 

		\hfill
		\rotatebox{90}{\parbox{3cm}{\centering Cu20W80}}\ 
		\subfigure[]{\includegraphics[width=3.7cm]{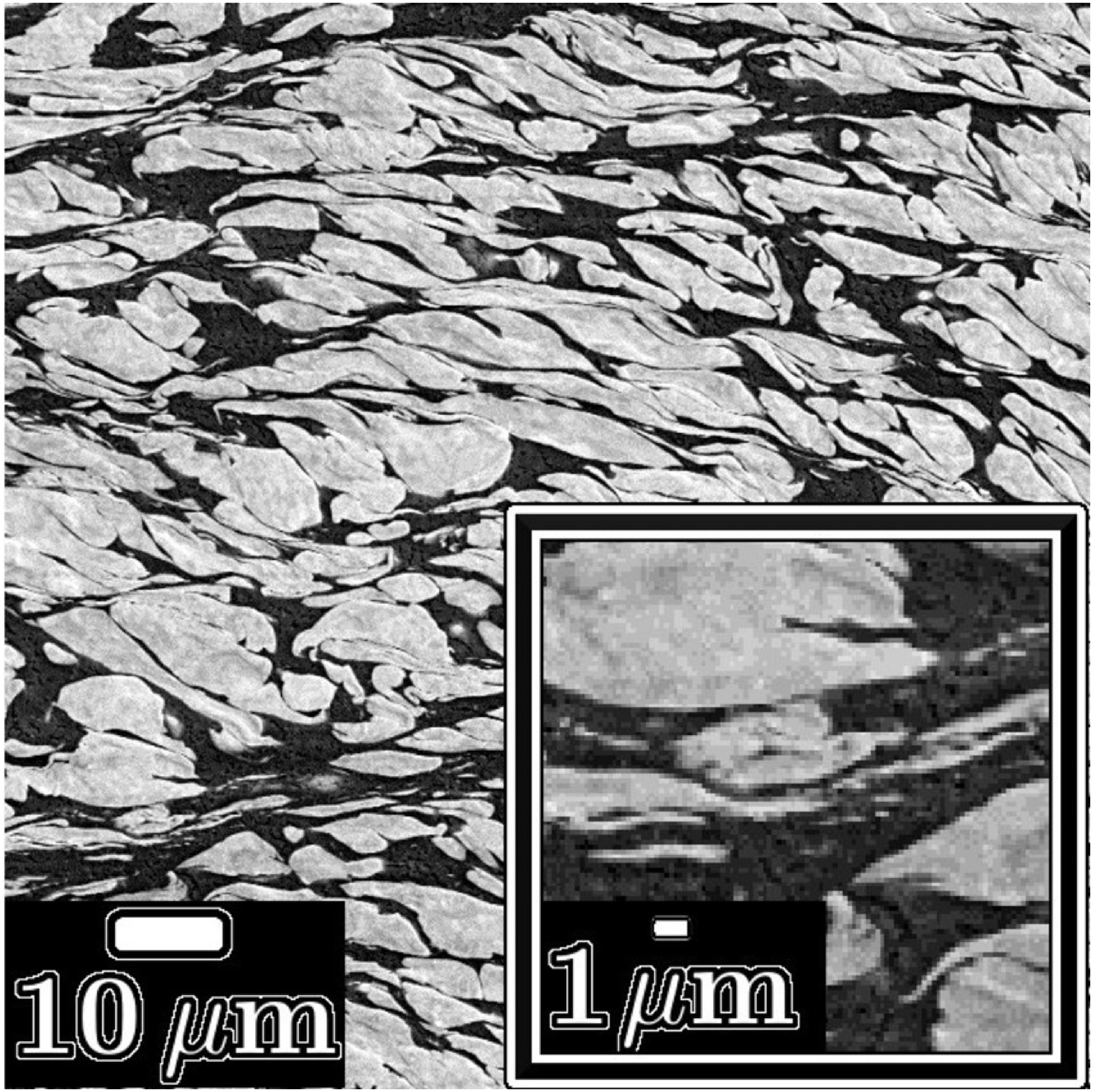}} \hfill 
		\subfigure[]{\includegraphics[height=3.7cm]{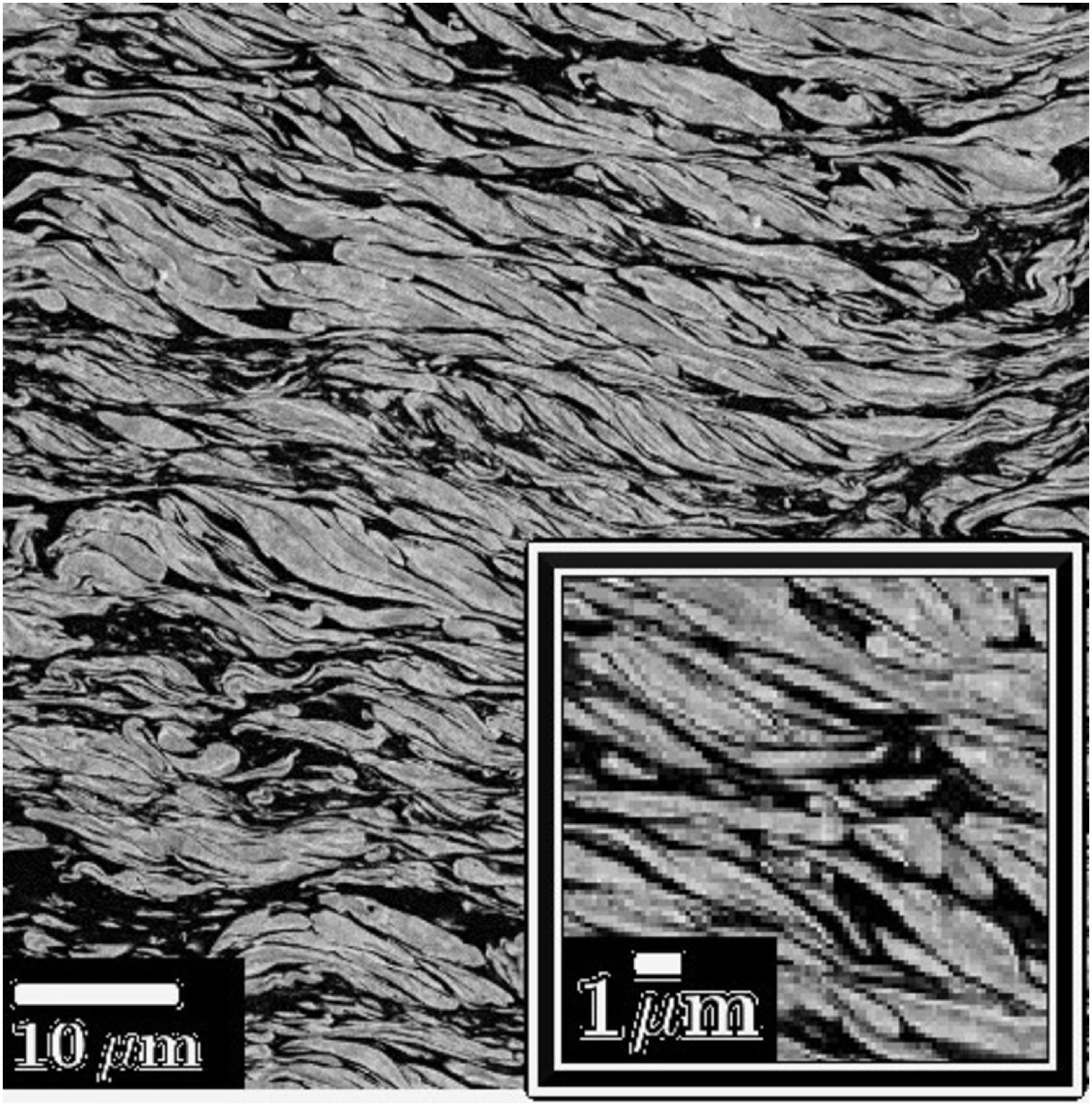}}\hfill\
 
		\caption{Backscattered electron micrographs of the microstructures of the Cu-refractory composites after step 1 HPT deformation, showing the near-centre region ($r\sim1$\,mm) (a,c,e,g), and edge ($r\sim14$\,mm)  (b,d,f,h). In each micrograph the shear plane is horizontal and normal to the plane of the page. Insets in each micrograph show the microstructure at higher magnification. \label{fig-sem-cu-all}}
	\end{center}
\end{figure}

All composites adopted a lamellar morphology after HPT deformation to a strain of $\sim$75. Figure~\ref{fig-sem-cu-all} shows representative backscattered electron images of each composite, with the  shear plane horizontal in each micrograph.  The refractory particles have undergone considerable plastic deformation, elongating in the shear direction, although the orientation of individual particles can vary substantially. The level of microstructural refinement is clearly greater at the rim of the sample (Figures~\ref{fig-sem-cu-all}(b,d,f,h)) than at the centre (a,c,e,g), due to the radial dependence of the equivalent strain. The microstructures of the Cu-W and Cu-Mo alloys are more uniform than that of Cu-Cr, in which much coarser Cr particles can be identified (See Fig~\ref{fig-sem-cu-all}(b)).

The refractory particle size  in Cu57Cr43 decreased substantially from 41\,$\mu$m in the as-received condition to  1.3\,$\mu$m (normal to the shear plane) after deformation (See Table~\ref{tab-stereo}). The HPT-deformed Cu50Mo50 and Cu30Mo70 had similar layer thickness values of 450\,nm and 400\,nm, respectively. Although the initial particle size in Cu20W80 was smaller than in the Mo-containing composites the HPT-deformed microstructure was coarser, with a layer thickness of 900\,nm.

\subsubsection{Texture Analysis\label{sssec-texture1}}

\begin{figure*}
	\begin{center}
		\begin{minipage}[b]{0.9\textwidth}
		\hspace*{1ex}\hfill Cu57Cr43\hfill\hfill Cu50Mo50 \hfill\hfill Cu30Mo70 \hfill\hfill Cu20W80\hfill\ 

	\rotatebox{90}{\parbox{3.5cm}{\centering 110}}
		\hfill
		\subfigure[($\alpha=11^\circ$)\label{fig-APF_Cu57Cu43_110}]{\includegraphics[width=3.4cm]{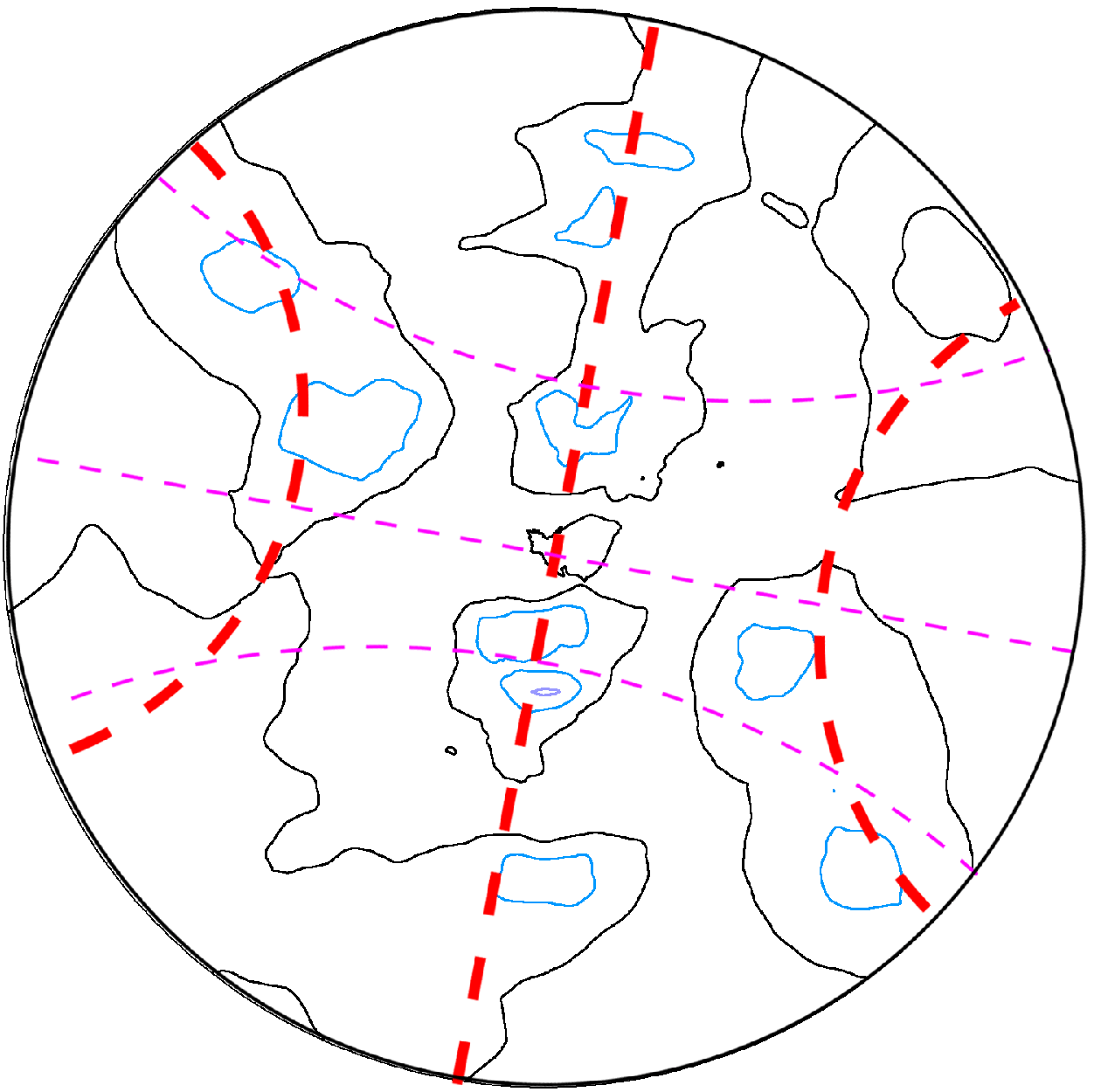}}
		\hfill
		\subfigure[\label{fig-APF_Cu50Mo50_110}($\alpha=12^\circ$)]{\includegraphics[width=3.4cm]{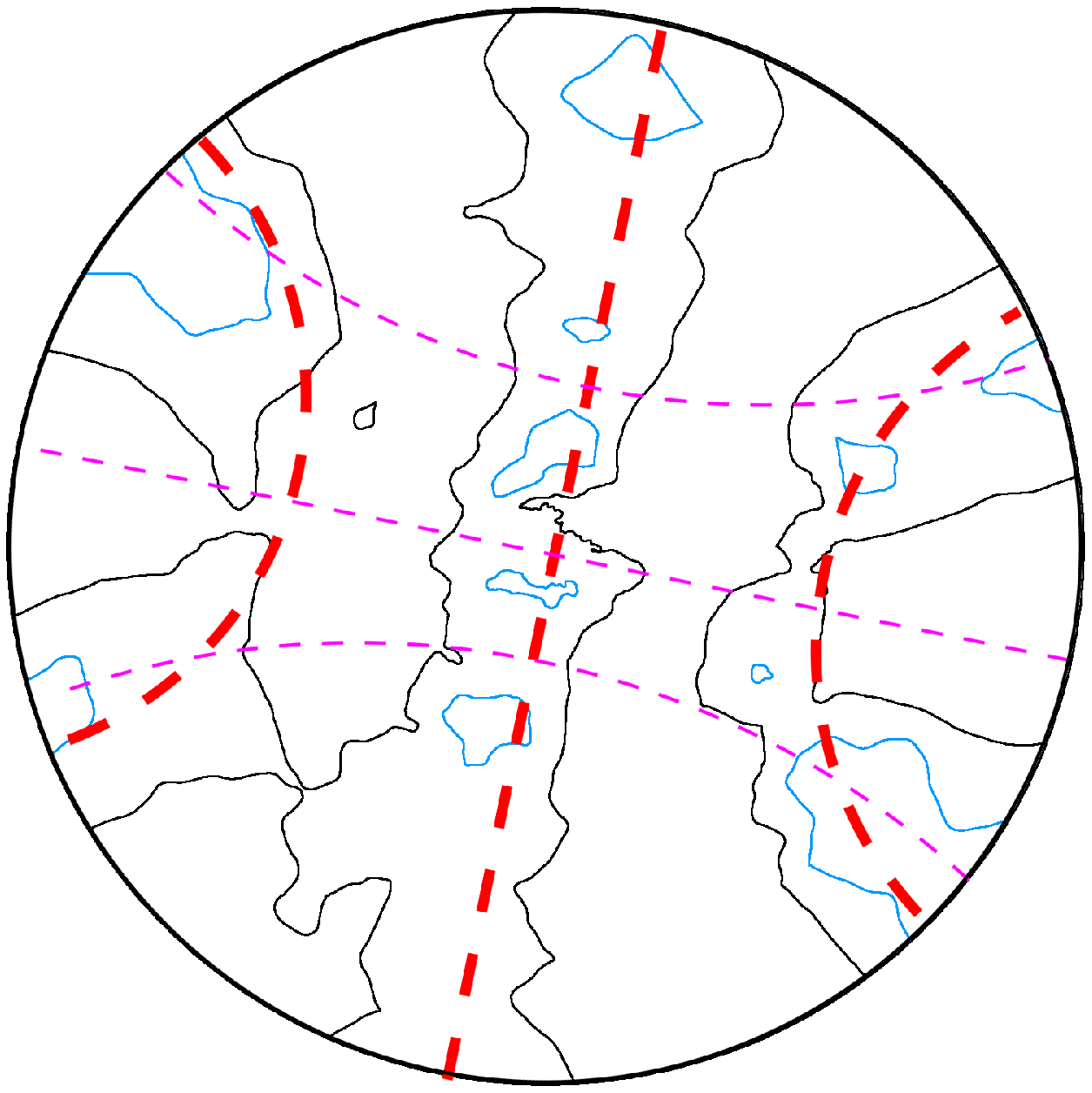}}
		\hfill
		\subfigure[\label{fig-APF_Cr_110}($\alpha=8^\circ$)]{	\includegraphics[width=3.4cm]{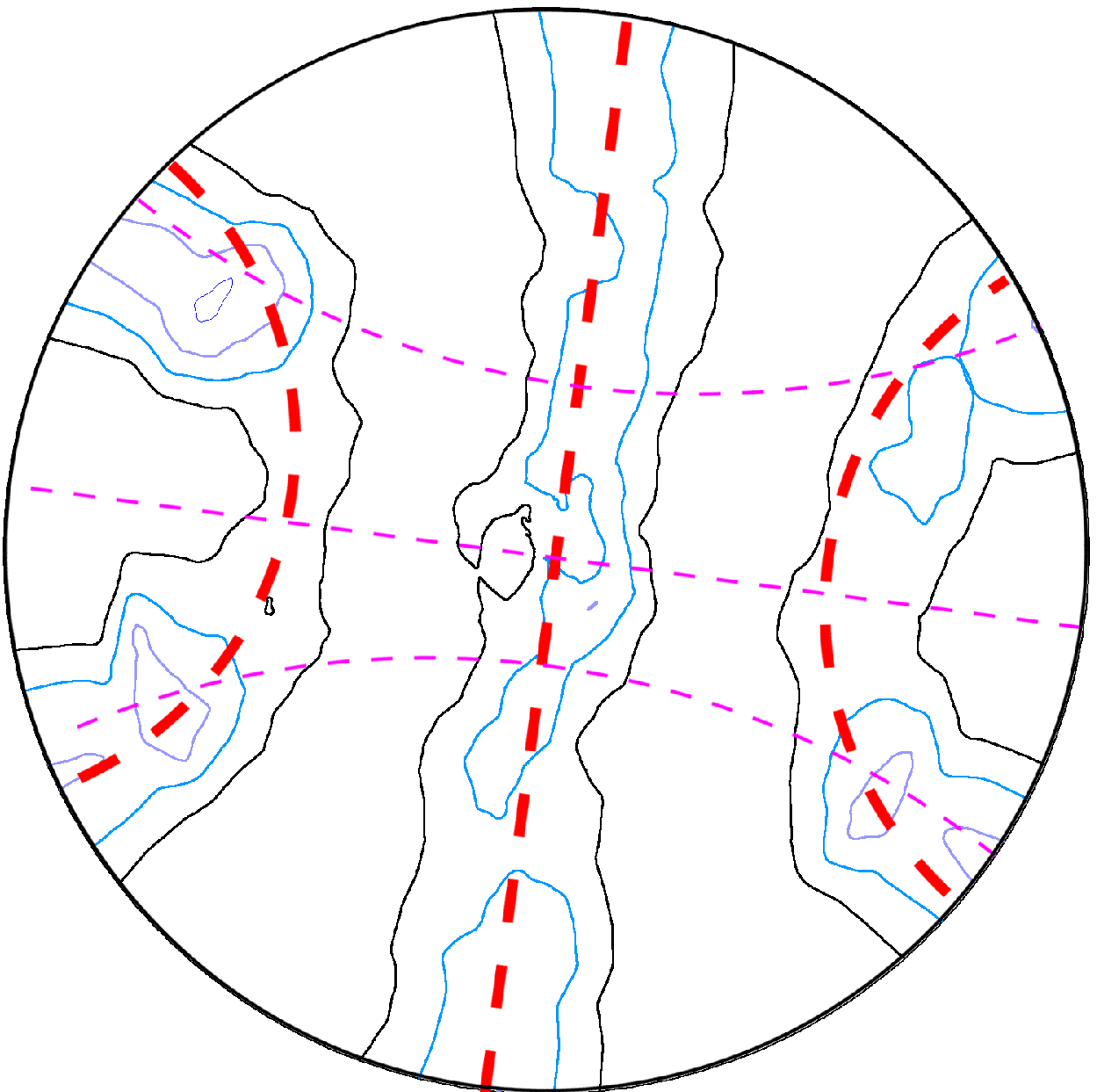}}
		\hfill
		\subfigure[\label{fig-APF_Cr_110}($\alpha=10^\circ$)]{	\includegraphics[width=3.4cm]{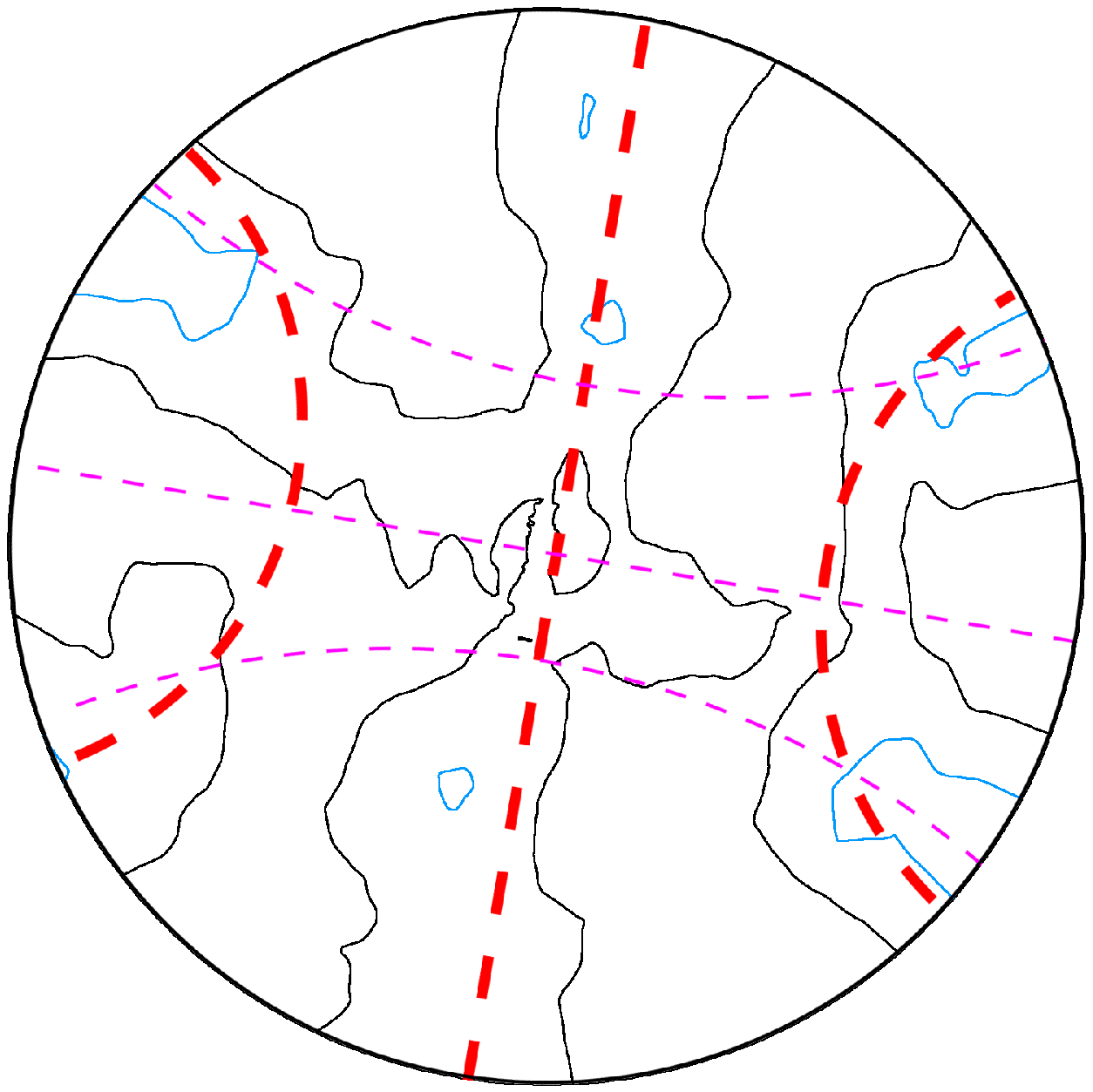}}
		\hfill

		\rotatebox{90}{\parbox{3.4cm}{\centering 222}}
		\hfill
		\subfigure[($\alpha=19^\circ$)\label{fig-APF_Cu57Cu43_222}]{\includegraphics[width=3.4cm]{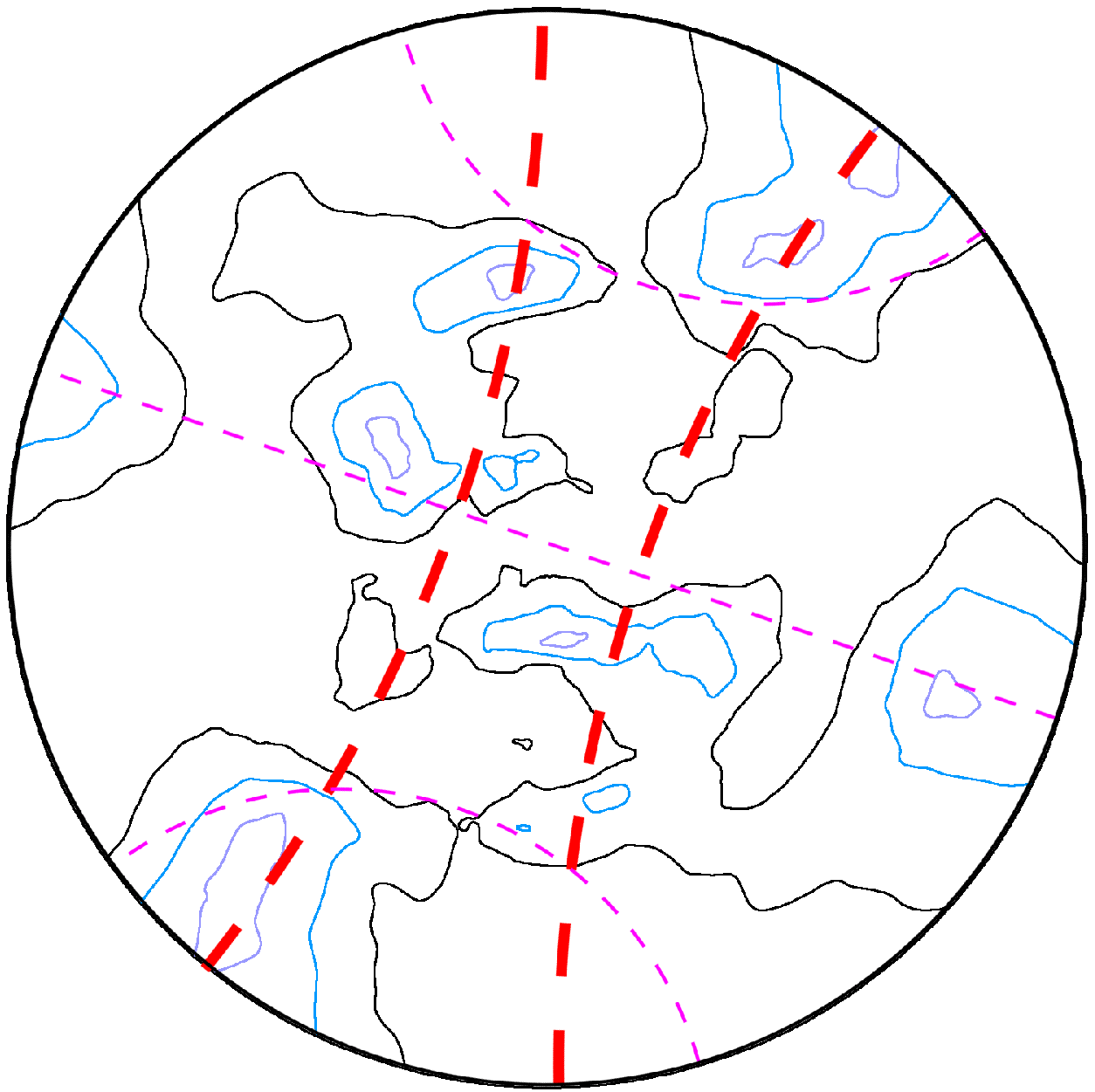}}
		\hfill
		\subfigure[\label{fig-APF_Cu50Mo50_222}($\alpha=11^\circ$)]{\includegraphics[width=3.4cm]{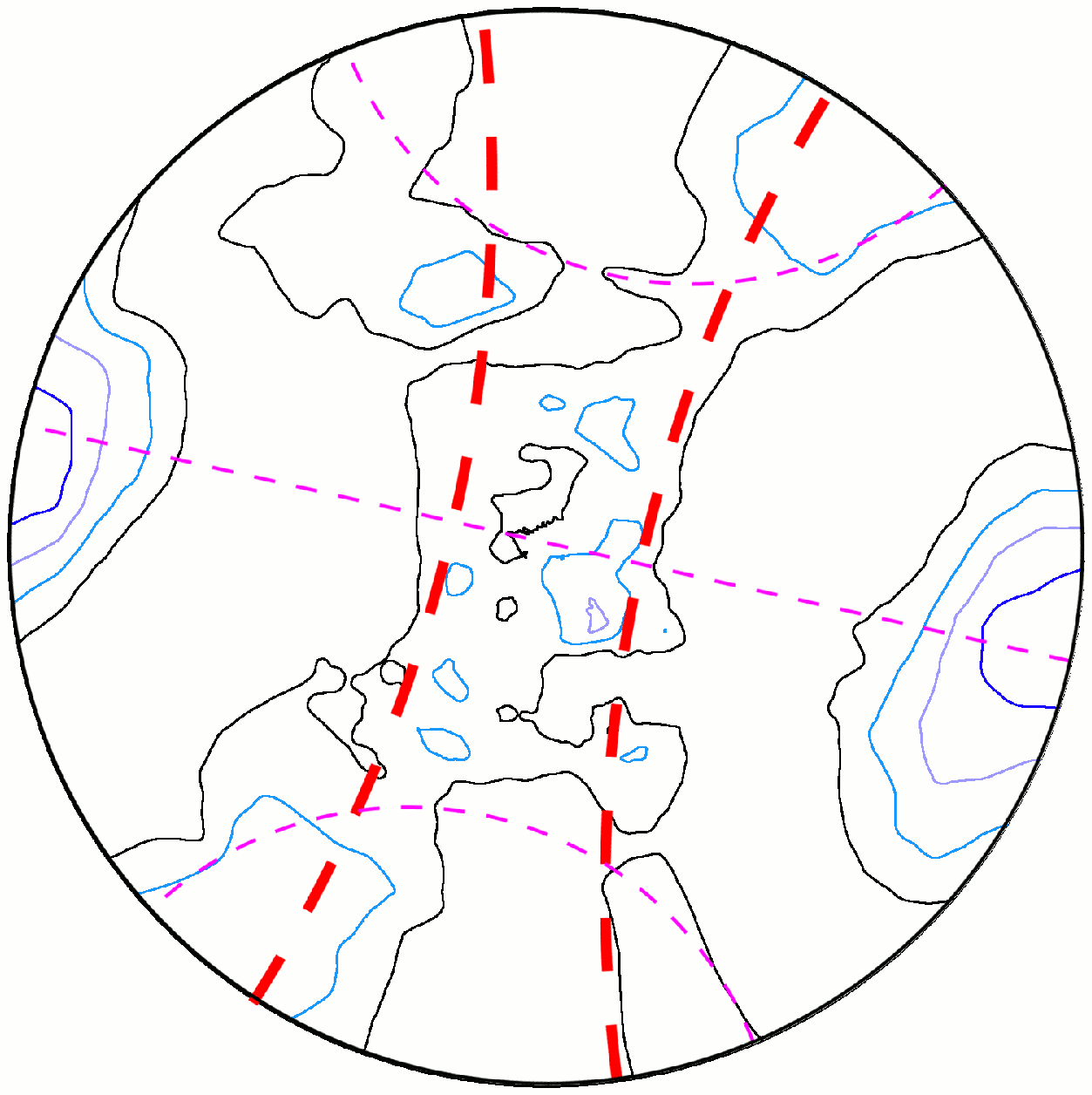}}
		\hfill
		\subfigure[\label{fig-APF_Cu30Mo70_222}($\alpha=8^\circ$)]{	\includegraphics[width=3.4cm]{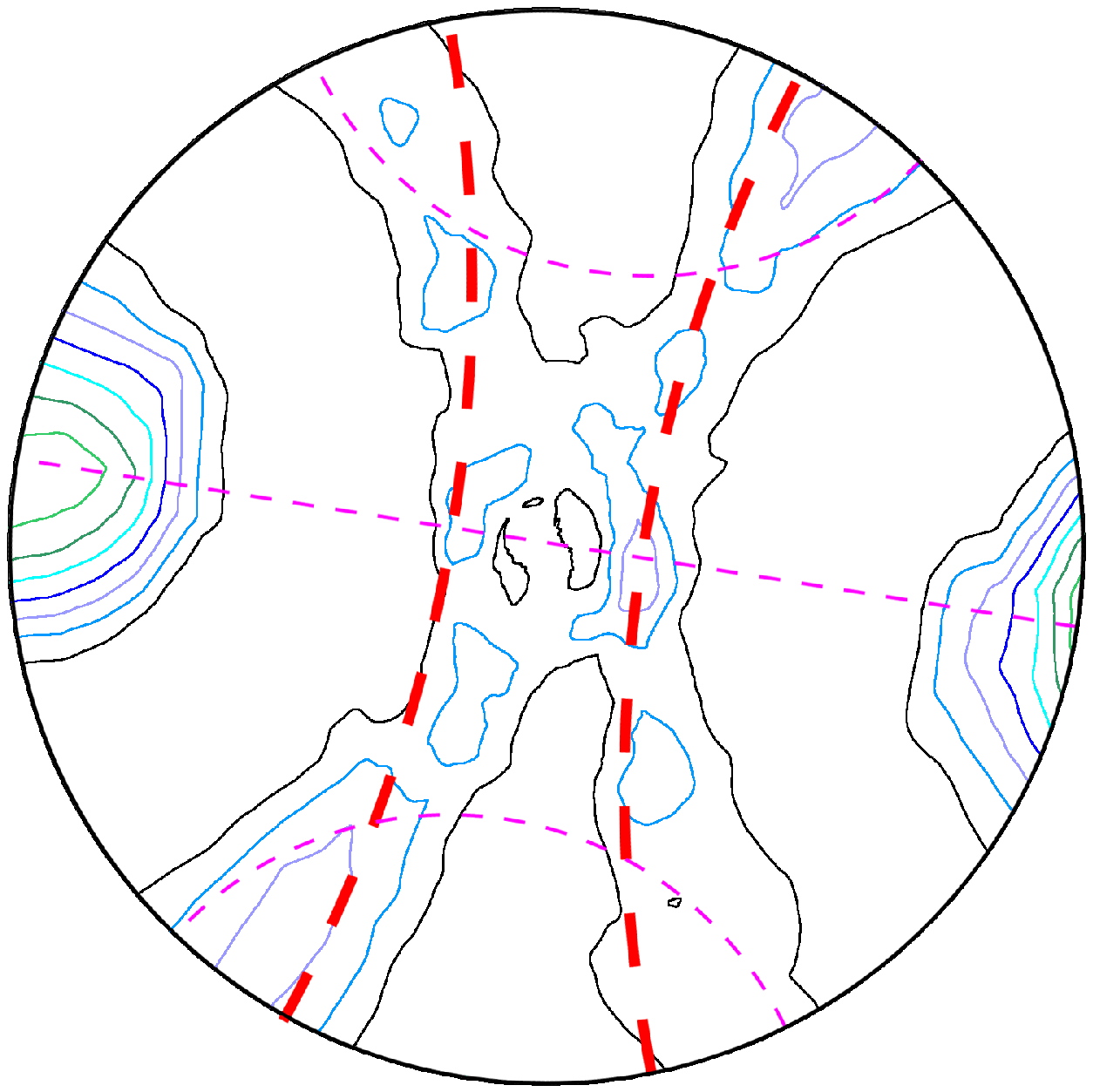}}
		\hfill
		\subfigure[($\alpha=14^\circ$)]{	\includegraphics[width=3.4cm]{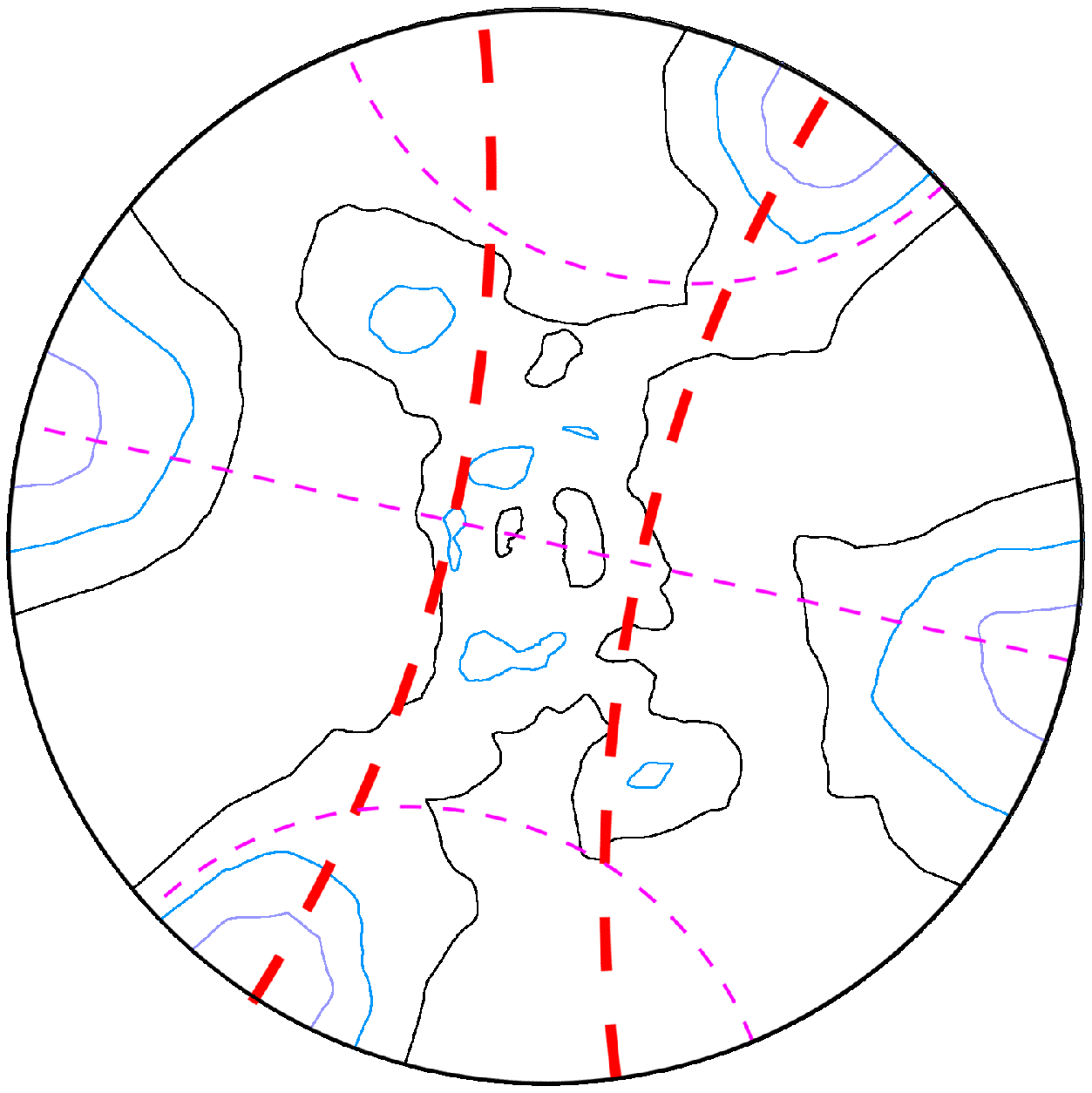}}
		\hfill\

		\end{minipage}
		\begin{minipage}[b]{0.07\textwidth}
		\includegraphics[width=0.9\textwidth]{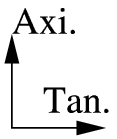} 		

		\includegraphics[width=0.9\textwidth]{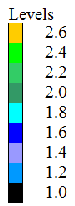} 
		\end{minipage} 
		\caption{Pole figures showing the texture of the refractory element after step 1 HPT deformation ($\epsilon_1\sim75$). Traces showing the ideal shear texture for BCC in (110) \cite{BaczynskiJonas1996} and (222) pole figures are overlaid on the experimental data. Long dashes (red in the colour version) indicate $\{hkl\}\langle111\rangle$ fibre textures (defined as $D_1$ and $D_2$ in \cite{BaczynskiJonas1996}). Short dashed lines (magenta in the colour version) indicate $\{110\}\langle uvw\rangle$ fibre textures ($J_1$ and $J_2$ (\textit{ibid})). The texture of the refractory elements correspond to a $\{hkl\}\langle111\rangle$ fibre texture, with a rotation ($\alpha$) around the normal direction. \label{fig-110-Mo-overlay}
>>>>>>> master
\label{fig-pf-refractory-stage1}}
	\end{center}
\end{figure*}

At an imposed equivalent strain of $\sim$75 the refractory components in Cu-Mo and Cu-W composites showed   $\{hkl\}\langle111\rangle$ fibre textures.  Recalculated (110) and (222) pole figures (PFs) for the refractory metal in each composite are provided in Figure~\ref{fig-pf-refractory-stage1}, with the ideal shear texture for a body-centred cubic (BCC) phase overlaid on the experimental data. In the (110) pole figures the fibre texture was most pronounced in Cu50Mo50 and Cu30Mo70 with a maximum intensity of 1.8 and 1.7 respectively. (All texture intensities are expressed as multiples of the intensity for a random distribution (MRD)).  The Cu57Cr43 composite had a maximum intensity of 1.4 on the 110 PF, similar to Cu-W, but with a less well-defined texture.  The fibres were tilted away from the ideal shear texture and measurements on the (110) PF yielded tilt angles of 8$^\circ$ in Cu30Mo70, 10$^\circ$ in Cu20W80 and 12$^\circ$ in Cu50Mo50 (The texture data is summarised in Table~\ref{tab-texture-strength} in the supplementary material). 

In the (222) pole figures  there is higher intensity in the tangential direction, particularly in Cu-Mo samples (See Figure~\ref{fig-pf-refractory-stage1}(e-h)). The tilt angles calculated from the (222) pole figures differed from those for (110) by less than the 5$^\circ$ step size used for the texture scans, except for Cu57Cr43, where the tilt was estimated as 11$^\circ$ from (110) and 19$^\circ$ from (222). 

Evaluating the texture of copper proved to be impractical. The low diffracted intensity of Cu compared to W made it impossible to determine the texture for Cu in Cu-W. Although there was sufficient diffracted intensity in the two Cu-Mo composites, measurements revealed only weak, and poorly defined textures. It was only in Cu57Cr43 that a moderate (max=2.3) $\{hkl\}\langle110\rangle$ texture  \cite{TothNeale1989,GhoshRenk2017} was observed. Further analysis of the texture was therefore restricted to the refractory metal texture, however, the (111) pole figure for Cu in Cu57Cr43 have been provided in Figure~\ref{fig-pf-cu57cu43_cu} of the supplementary material.

\subsection{HPT deformed (Second step)}

\subsubsection{Microstructure}

\begin{figure*}
	\begin{center}
	\hfill
	\subfigure[Cu57Cr43 ($\epsilon_2\sim1,500$) \label{fig-stage-two-stem-Cu57Cr43}]{\includegraphics[width=5cm]{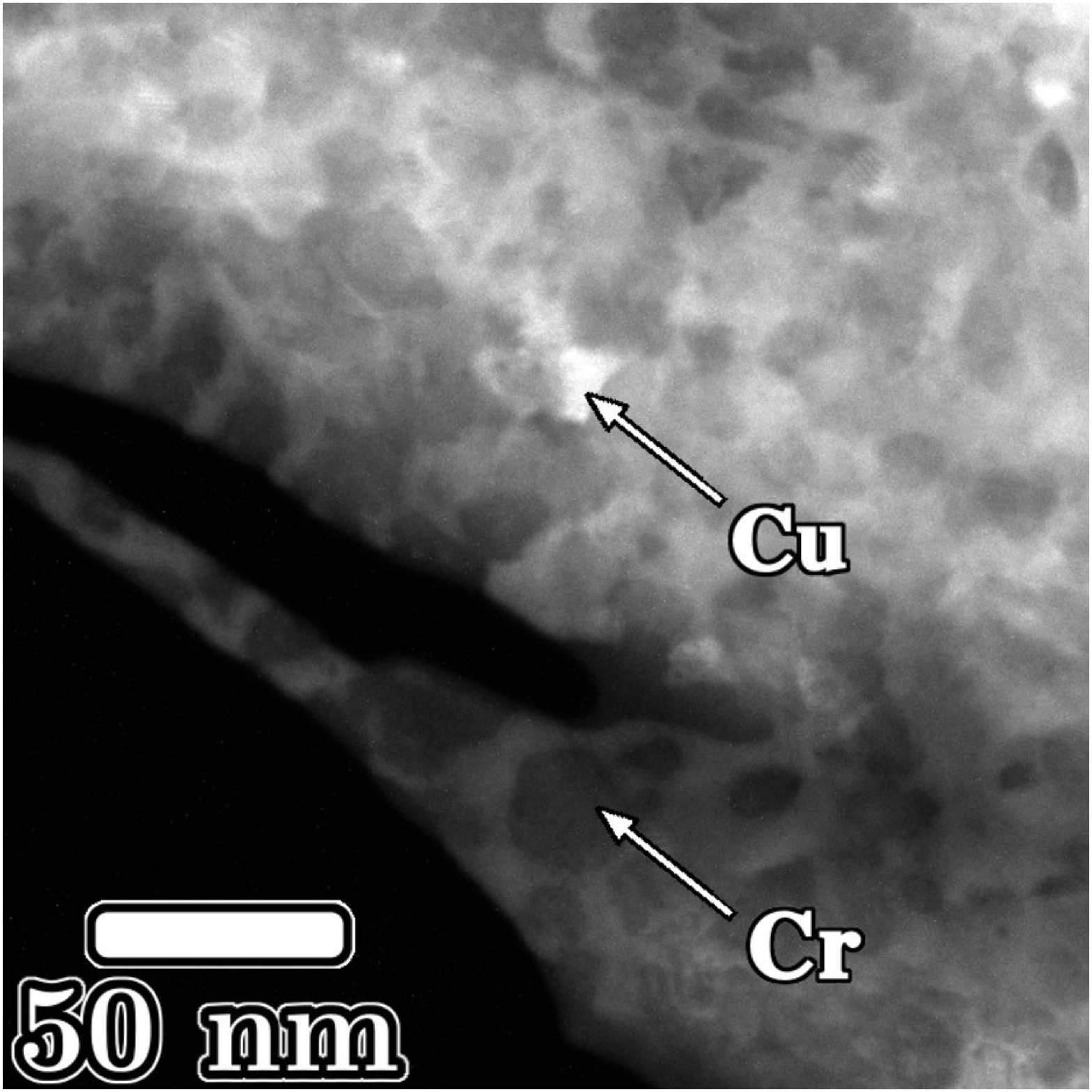}}
	\hfill
	\subfigure[Cu50Mo50 ($\epsilon_2\sim1,000$) \label{fig-stage-two-stem-Cu50Mo50}]{\includegraphics[width=5cm]{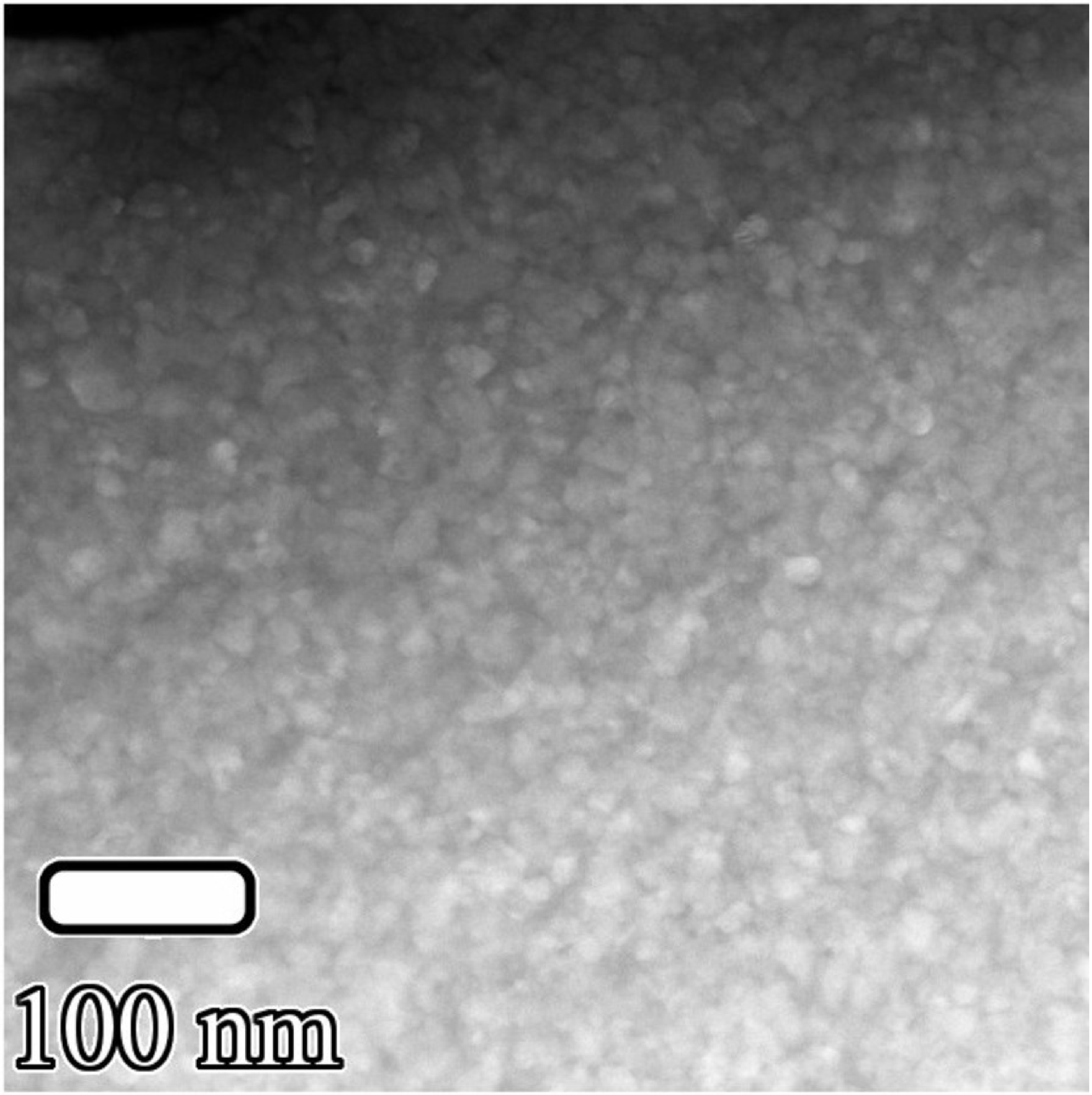}}
	\hfill\ 

	\hfill
	\subfigure[Cu30Mo70 ($\epsilon_2\sim1,000$) \label{fig-stage-two-stem-Cu30Mo70}]{\includegraphics[width=5cm]{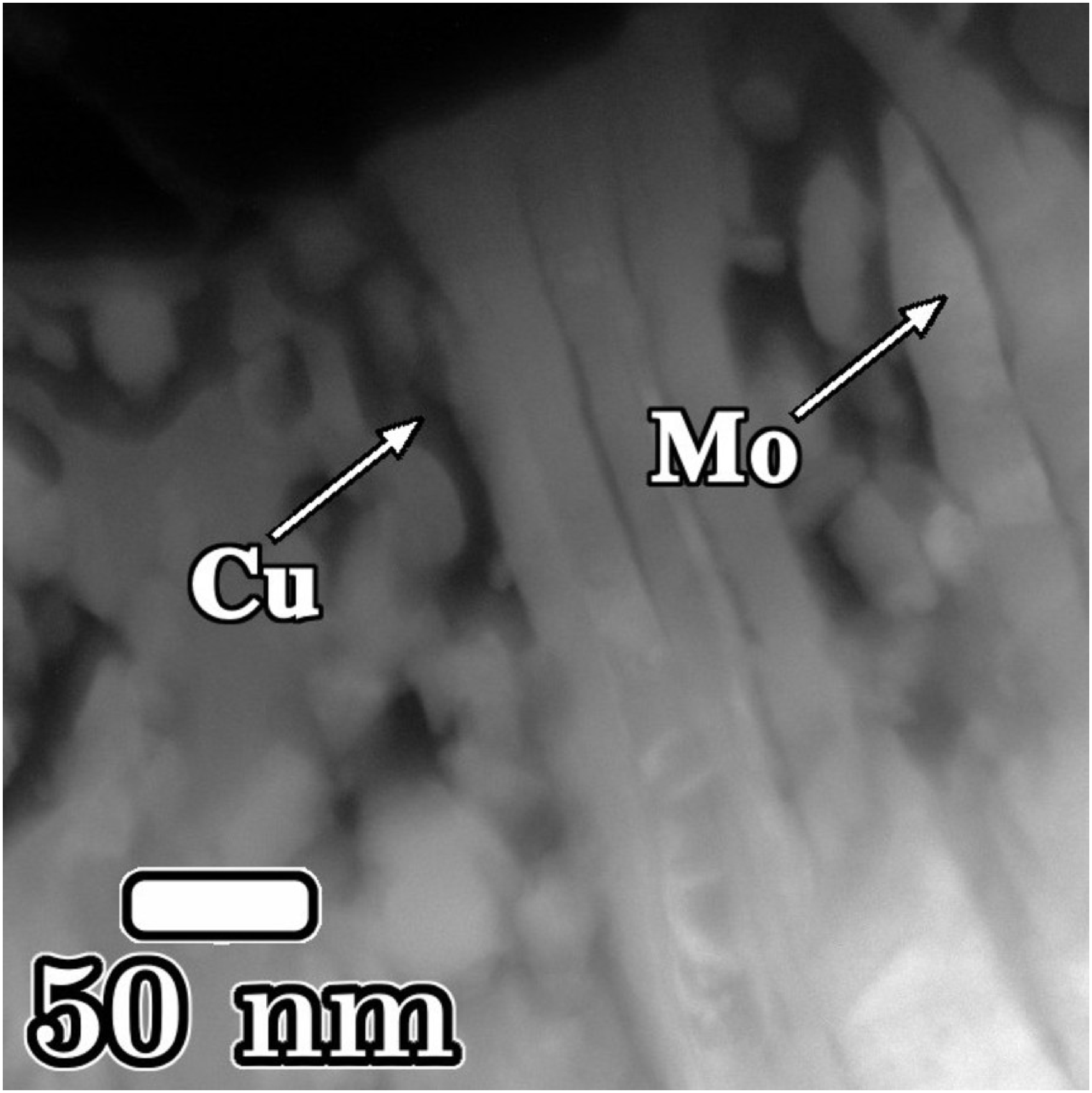}}
	\hfill
	\subfigure[Cu20W80 ($\epsilon_2\sim700$) \label{fig-stage-two-stem-Cu20W80}]{\includegraphics[width=5cm]{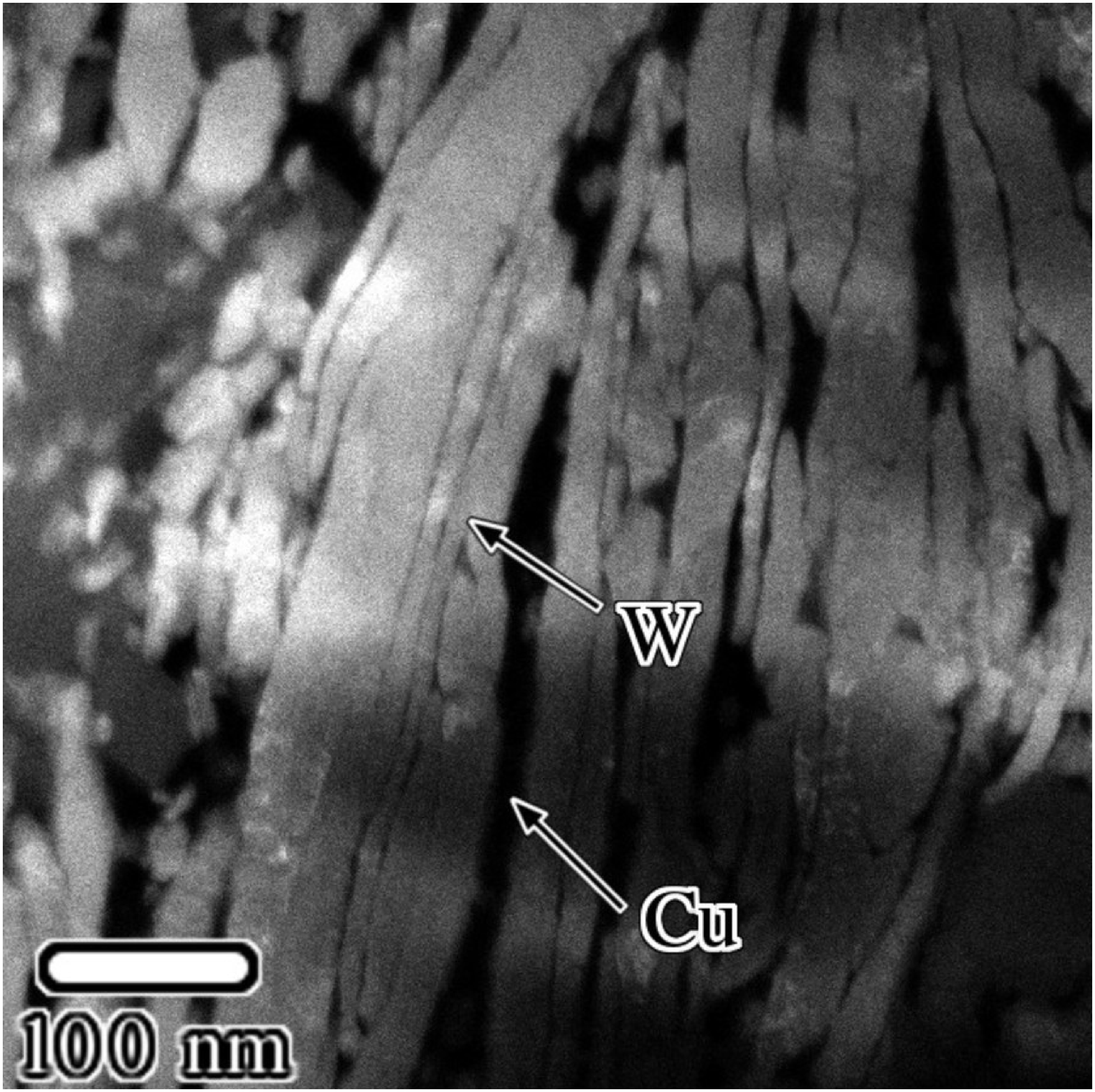}}
	\hfill\

	\caption{HAADF-STEM micrographs of the refractory composites after second step HPT deformation. The viewing direction is normal to the shear plane. In (c,d) the shear direction is approximately vertical (in the plane of the page).  \label{fig-stage-two-stem}}
	\end{center}
\end{figure*}

Significant differences between the composites became evident after second-step HPT deformation. Figure~\ref{fig-stage-two-stem} presents typical high-angle annular dark field scanning TEM (HAADF-STEM) micrographs, in which the Cu and refractory components can be readily identified due to atomic number contrast. The lamellae decomposed in Cu57Cr43 and Cu50Mo50 leaving a microstructure consisting of equiaxed nanoscale grains, reported for each of these materials \cite{GuoRosalie2017,RosalieGuo2017}. The refractory particle size was 14\,nm in Cu57Cr43 and 17\,nm in Cu50Mo50. In contrast, Cu20W80 and Cu30Mo70 retained lamellar microstructures after 50 and 35 rotations of step 2 deformation, respectively, during which the refractory layer width was reduced to tens of nanometers. Lamellar spacings of 10--20\,nm and 20--50\,nm were measured for Mo and W, respectively. Cu was present between the refractory lamellae, often  as layers of $\sim5$\,nm   thickness. This type of microstructure has been shown for Cu30Mo70 \cite{RosalieGuo2017} but to the best of the authors knowledge is the first report of such nanolamellar Cu-W composites. 

\begin{figure*}
	\begin{center}
	\hfill
	\subfigure[Cu57Cr43 ($\epsilon_2\sim1,500$)\label{fig-stage-two-sad-Cu57Cr43}]{\includegraphics[width=6cm]{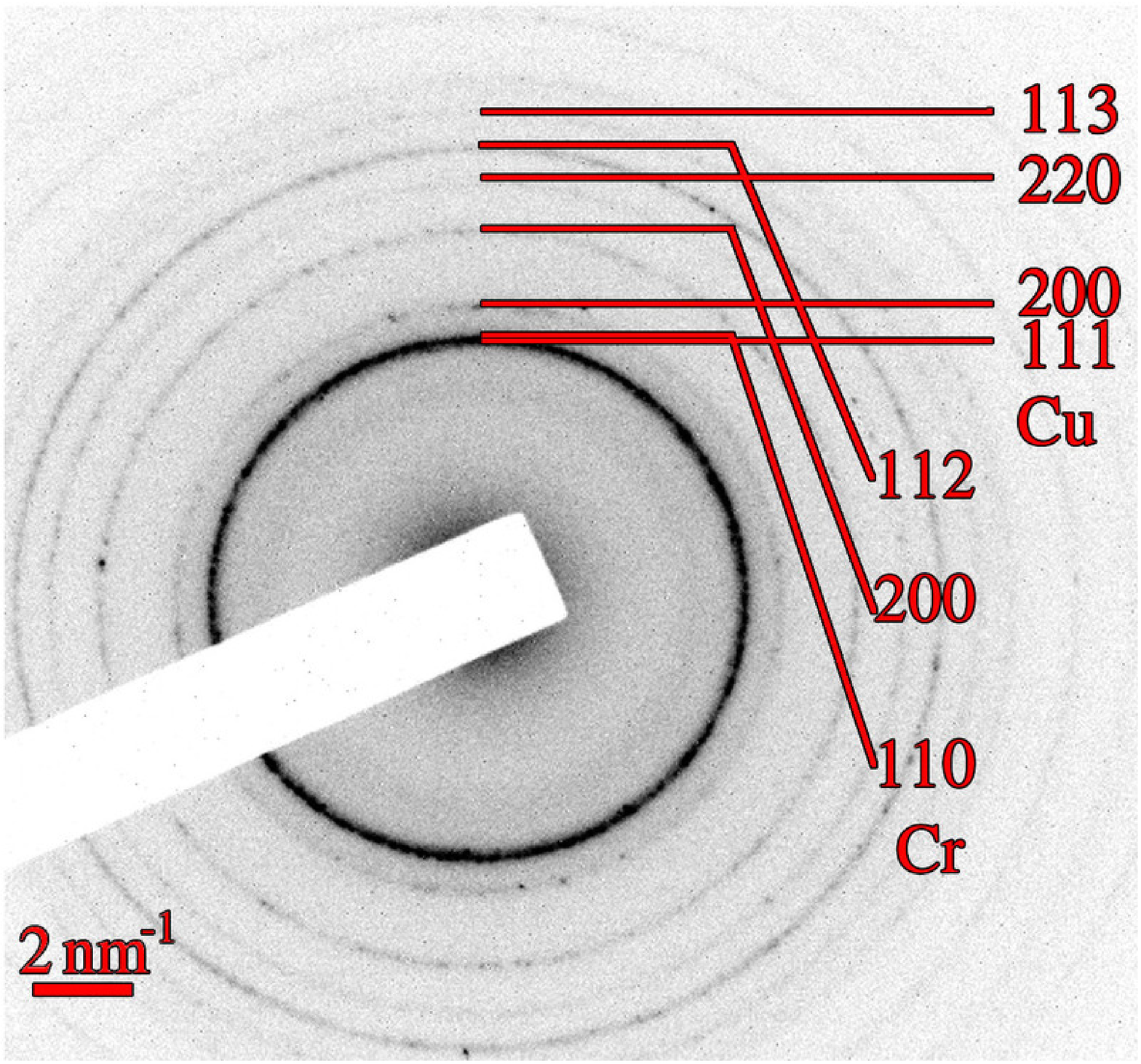}}
	\hfill
	\subfigure[Cu50Mo50 ($\epsilon_2\sim1,000$)\label{fig-stage-two-sad-Cu50Mo50}]{\includegraphics[width=6cm]{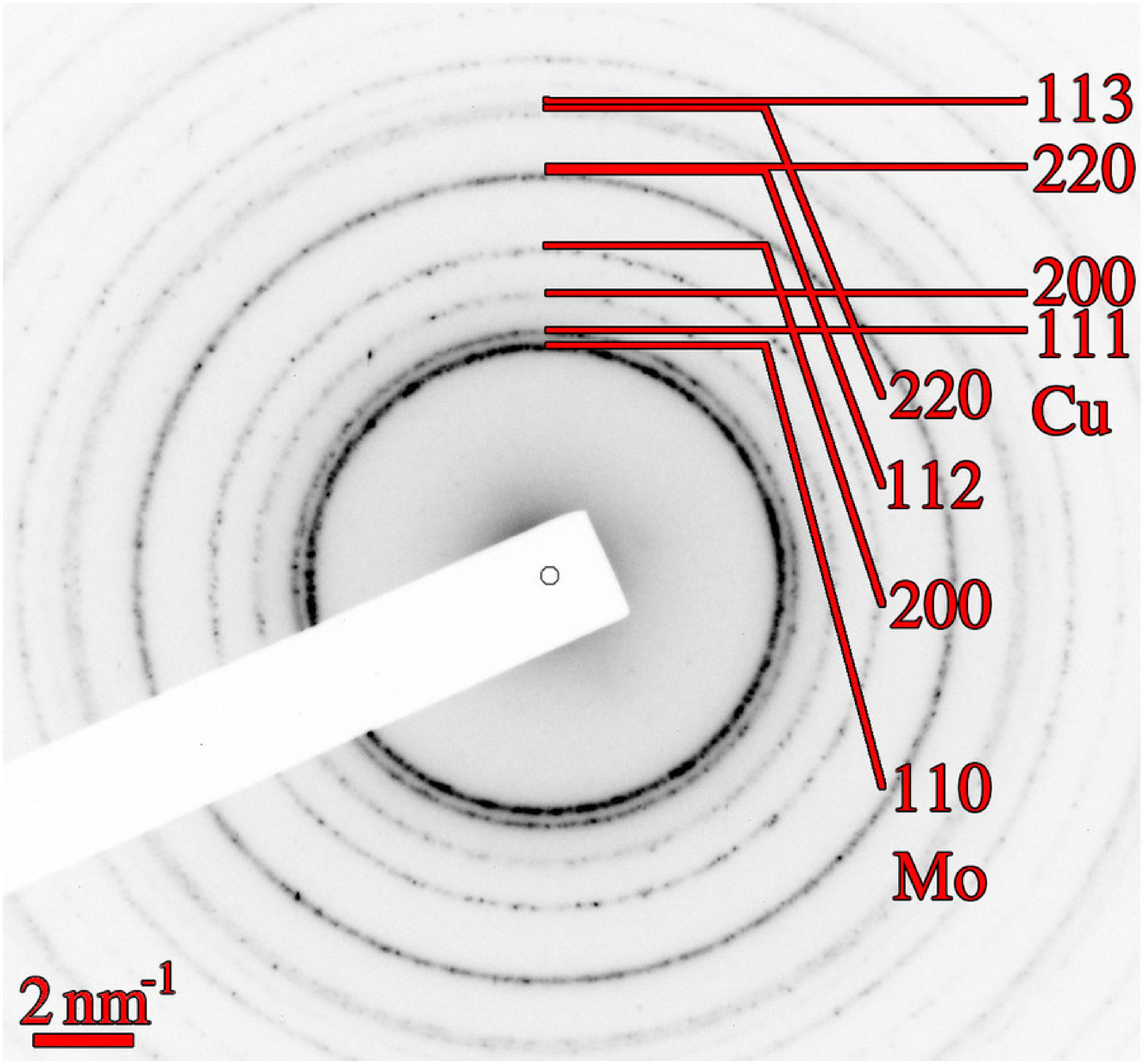}}
	\hfill\ 

	\hfill 
	\subfigure[Cu30Mo70 ($\epsilon_2\sim1,000$)\label{fig-stage-two-sad-Cu30Mo70}]{\includegraphics[width=6cm]{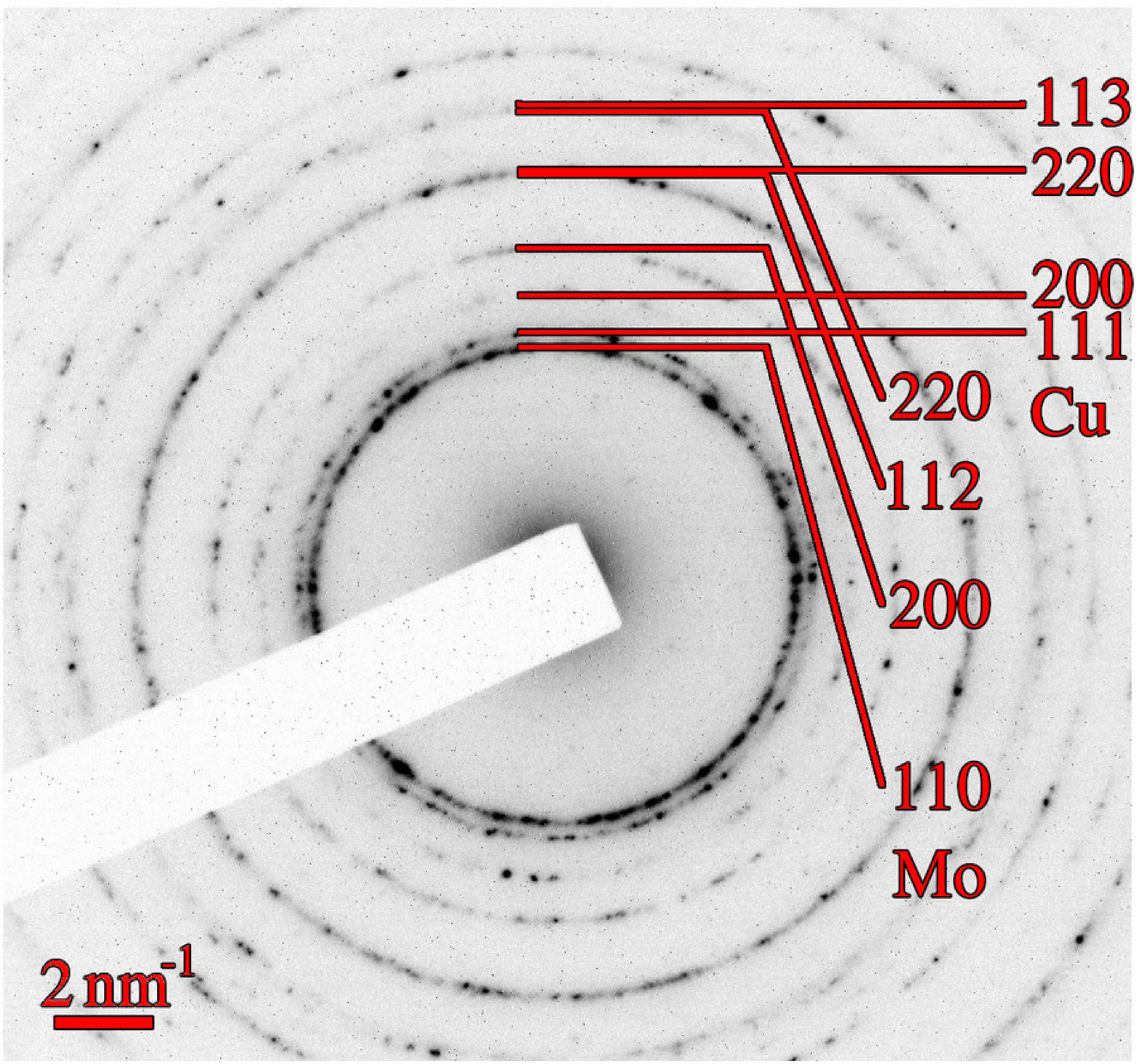}}
	\hfill
	\subfigure[Cu20W80 ($\epsilon_2\sim700$)\label{fig-stage-two-sad-Cu20W80}]{\includegraphics[width=6cm]{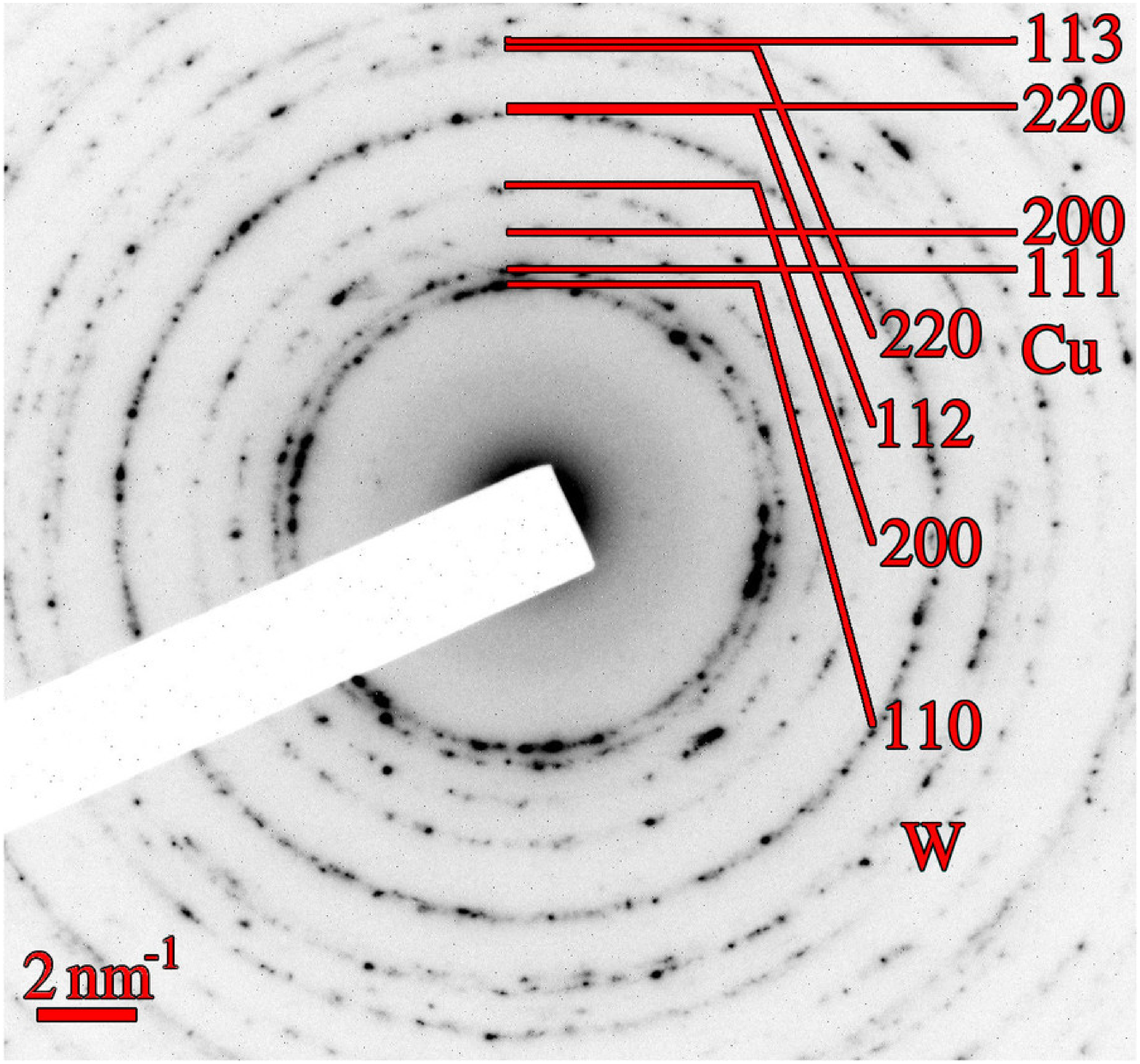}}
	\hfill\ 
	\caption{Selected area diffraction patterns obtained from each composites after second step HPT deformation. An identical aperture diameter of 120\,$\mu$m was used in each instance, and images were obtained close to the edge of the foil. The persistence of the lamellar microstructure is associated with local texture, as can be seen in (c,d). \label{fig-stage-two-sad}}
	\end{center}
\end{figure*}

Selected area diffraction (SAD) of the Cu57Cr43 (Fig.~\ref{fig-stage-two-sad-Cu57Cr43}) and Cu50Mo50 samples (Fig.~\ref{fig-stage-two-sad-Cu50Mo50}) yielded ring patterns with minimal radial variation in intensity. Composites which retained a lamellar microstructure displayed substantial radial variations in diffracted intensity.  This phenomenon indicates that the persistence of the lamellar structure is also associated with texture at a local scale, and is more pronounced in Cu20W80  (Fig.~\ref{fig-stage-two-sad-Cu20W80}) than in Cu30Mo70  (Fig.~\ref{fig-stage-two-sad-Cu30Mo70}). Each of the SAD patterns was obtained close to the edge of the foil, using an identical aperture diameter of 120\,$\mu$m\footnote{Since the microscope was equipped with a spherical aberration ($C_s$) corrector, the aperture size can be regarded as a good measure of the actual area investigated.}.

\subsubsection{Texture Analysis}

\begin{figure*}
	\begin{center}
		\begin{minipage}[b]{0.9\textwidth}
		\hspace*{1ex}\hfill Cu57Cr43\hfill\hfill Cu50Mo50 \hfill\hfill Cu30Mo70 \hfill\hfill Cu20W80\hfill\ 

		\rotatebox{90}{\parbox{3.4cm}{\centering 110}}
		\hfill
		\subfigure[\label{fig-APF_Cu57Cr43_110-stage2}]{	\includegraphics[width=3.4cm]{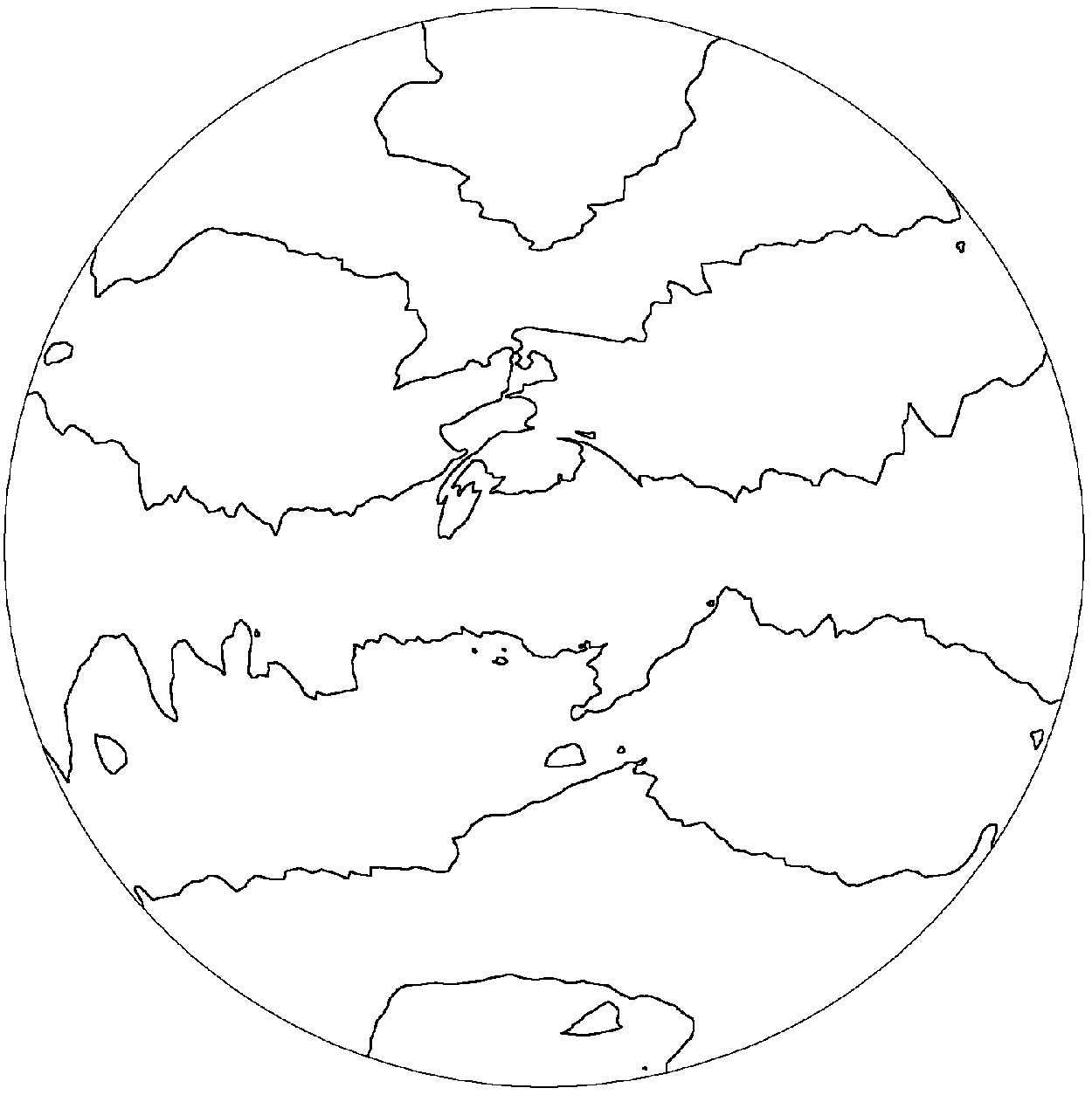}}
		\hfill
		\subfigure[\label{fig-APF_Cu50Mo50_110-stage2}]{	\includegraphics[angle=90,width=3.4cm]{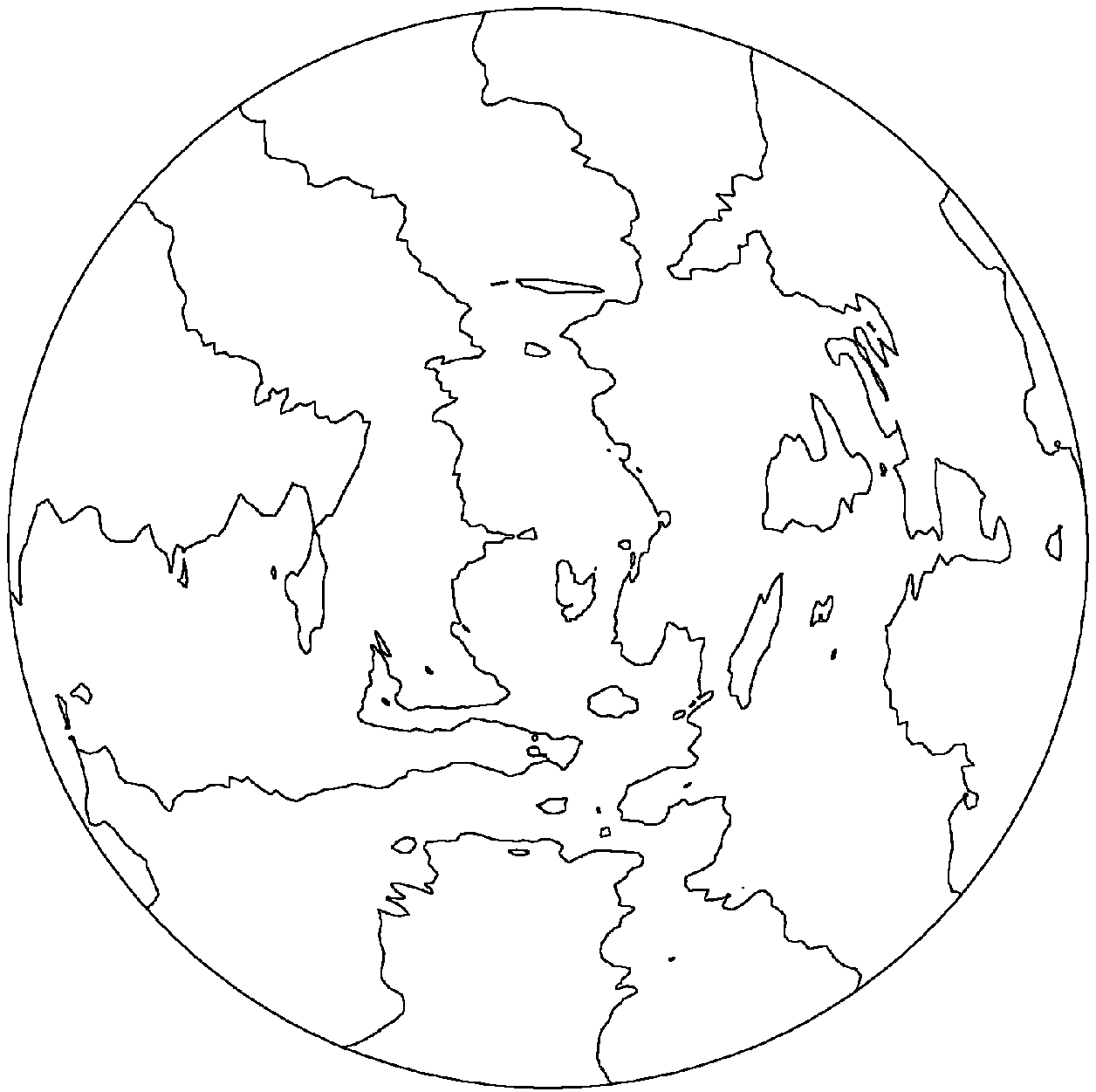}}
		\hfill
		\subfigure[($\alpha=-14^\circ$)\label{fig-APF_Cu30Mo70_110-stage2}]{	\includegraphics[width=3.4cm]{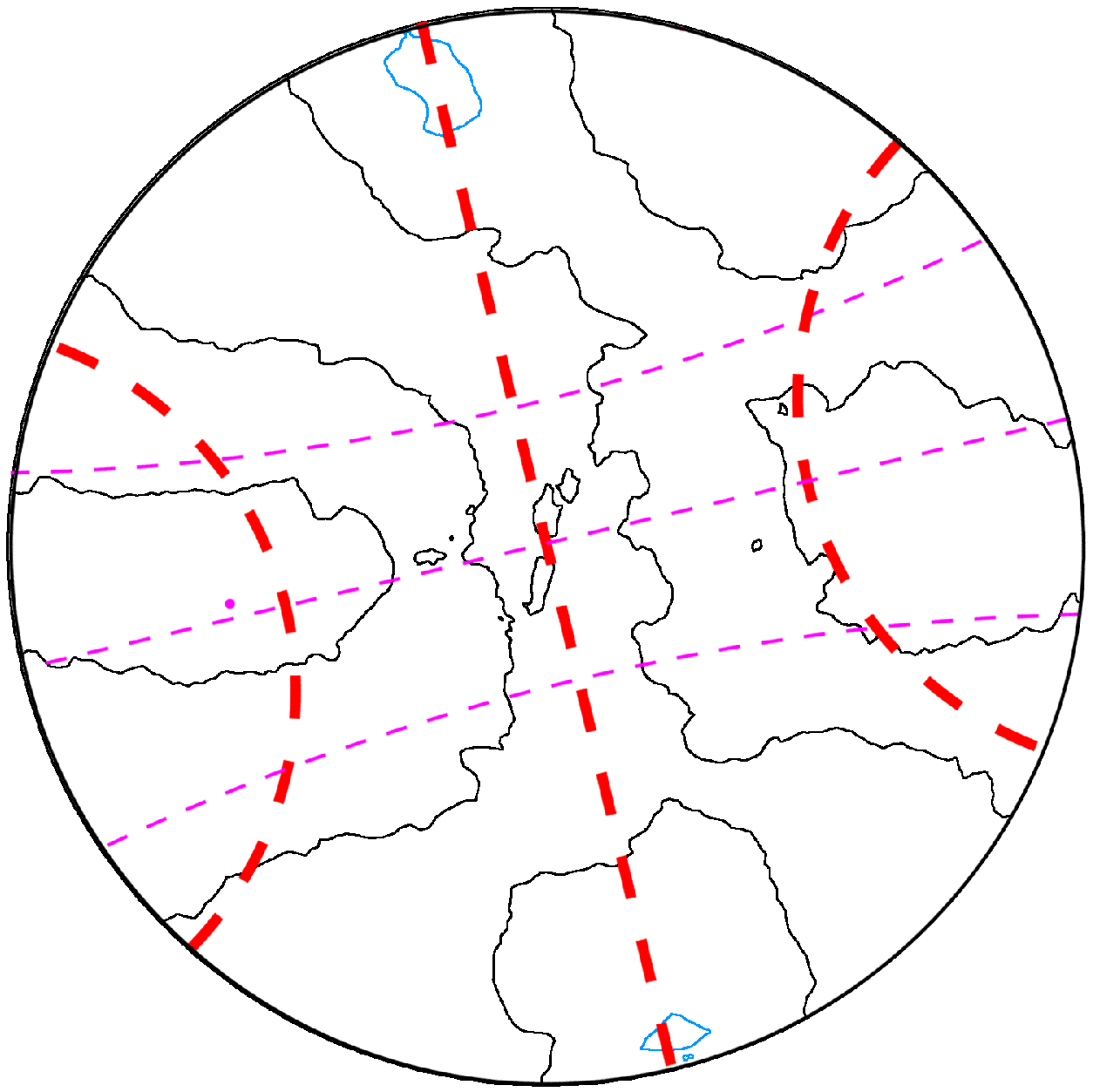}}
		\hfill
		\subfigure[($\alpha=0^\circ$)\label{fig-APF_Cu20W80_110-stage2}]{	\includegraphics[width=3.4cm]{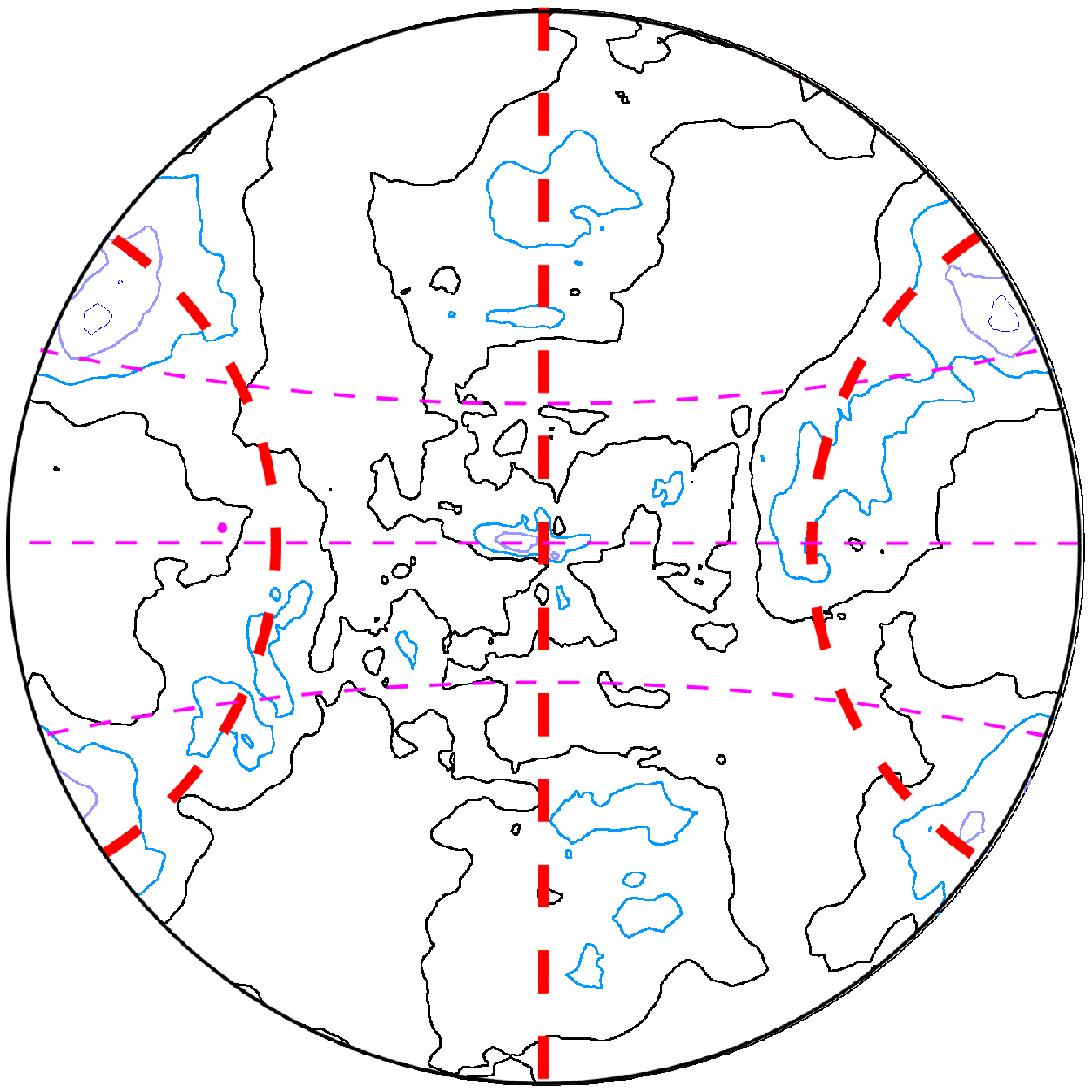}}
		\hfill

		\rotatebox{90}{\parbox{3.4cm}{\centering 222}}
		\subfigure[\label{fig-APF_Cu57Cr43_222-stage2}]{	\includegraphics[width=3.4cm]{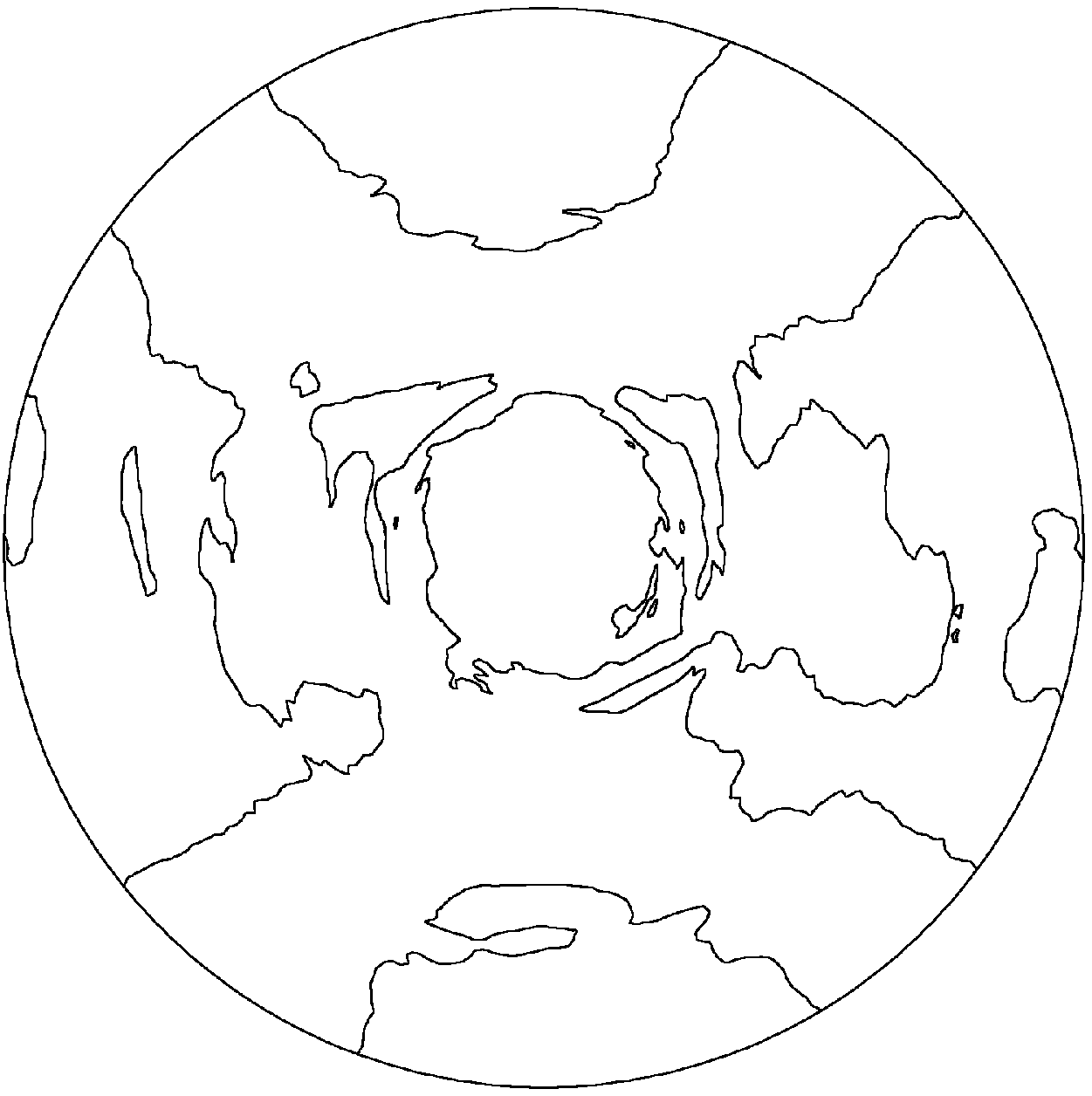}}\hfill
		\hfill
		\subfigure[\label{fig-APF_Cu50Mo50_222-stage2}]{\includegraphics[angle=90,width=3.4cm]{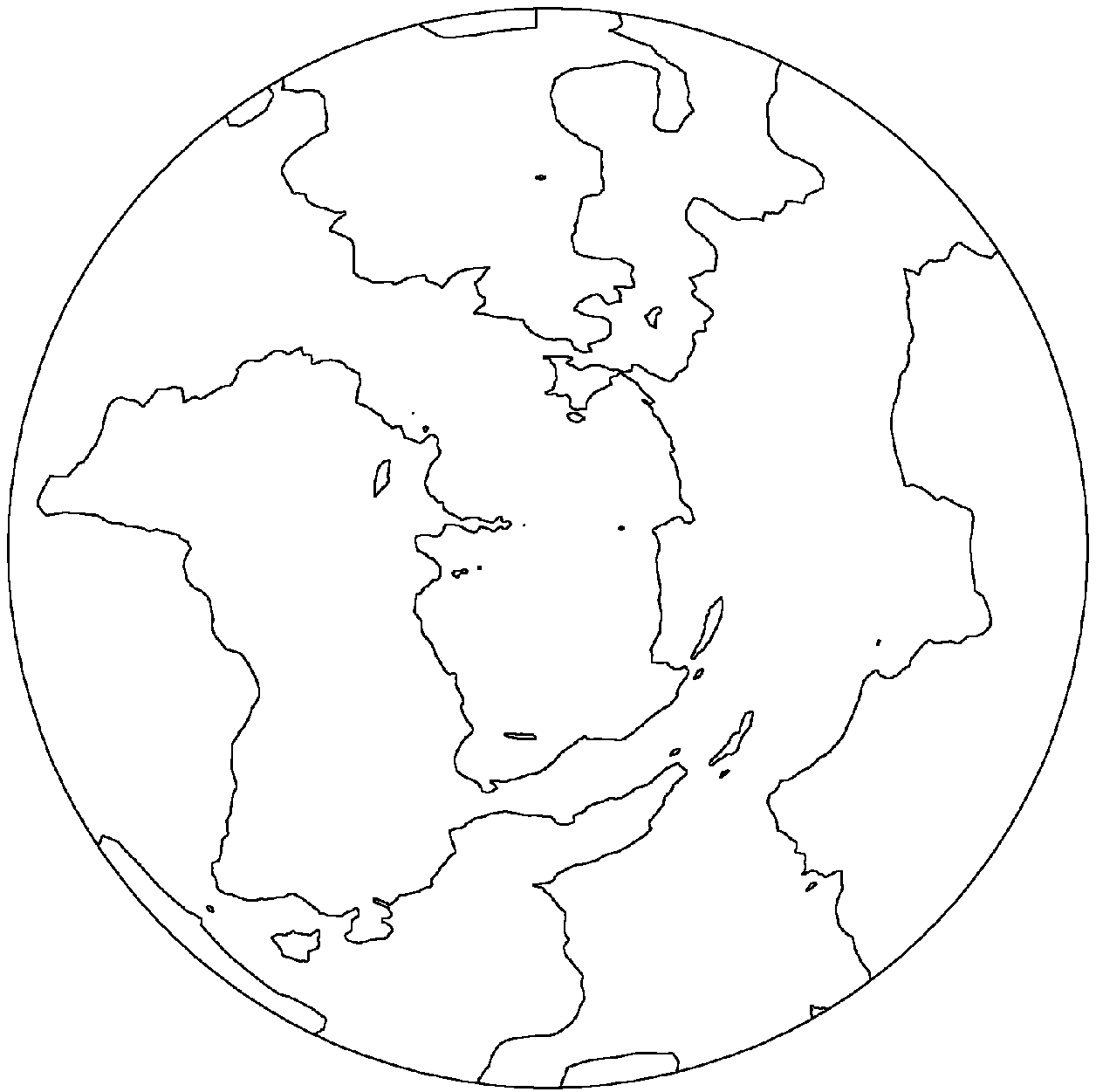}}\hfill
		\hfill
		\subfigure[($\alpha=-14^\circ$)\label{fig-APF_Cu30Mo70_222-stage2}]{	\includegraphics[width=3.4cm]{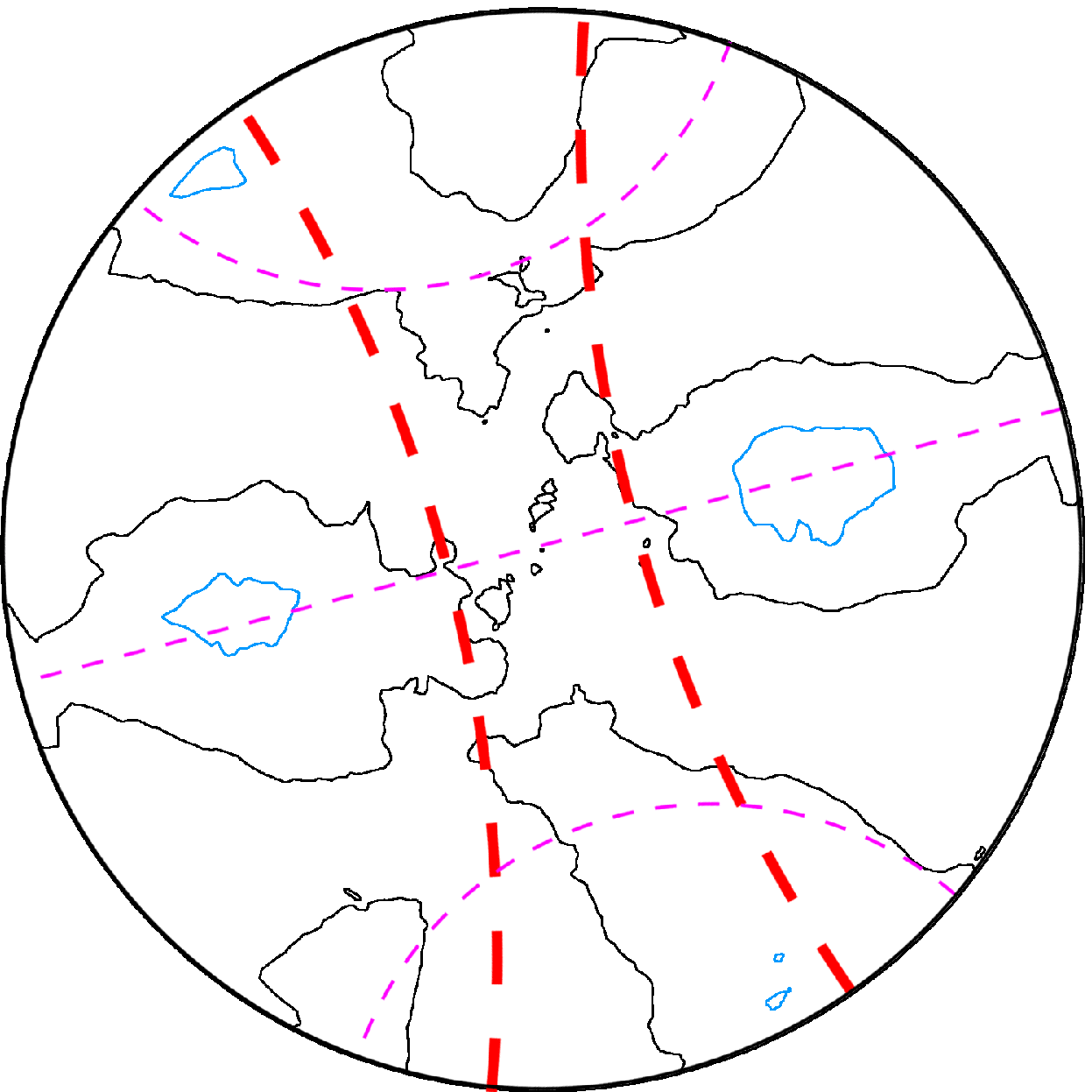}}
		\hfill
		\subfigure[($\alpha=0^\circ$)\label{fig-APF_Cu20W80_222-stage2}]{	\includegraphics[width=3.4cm]{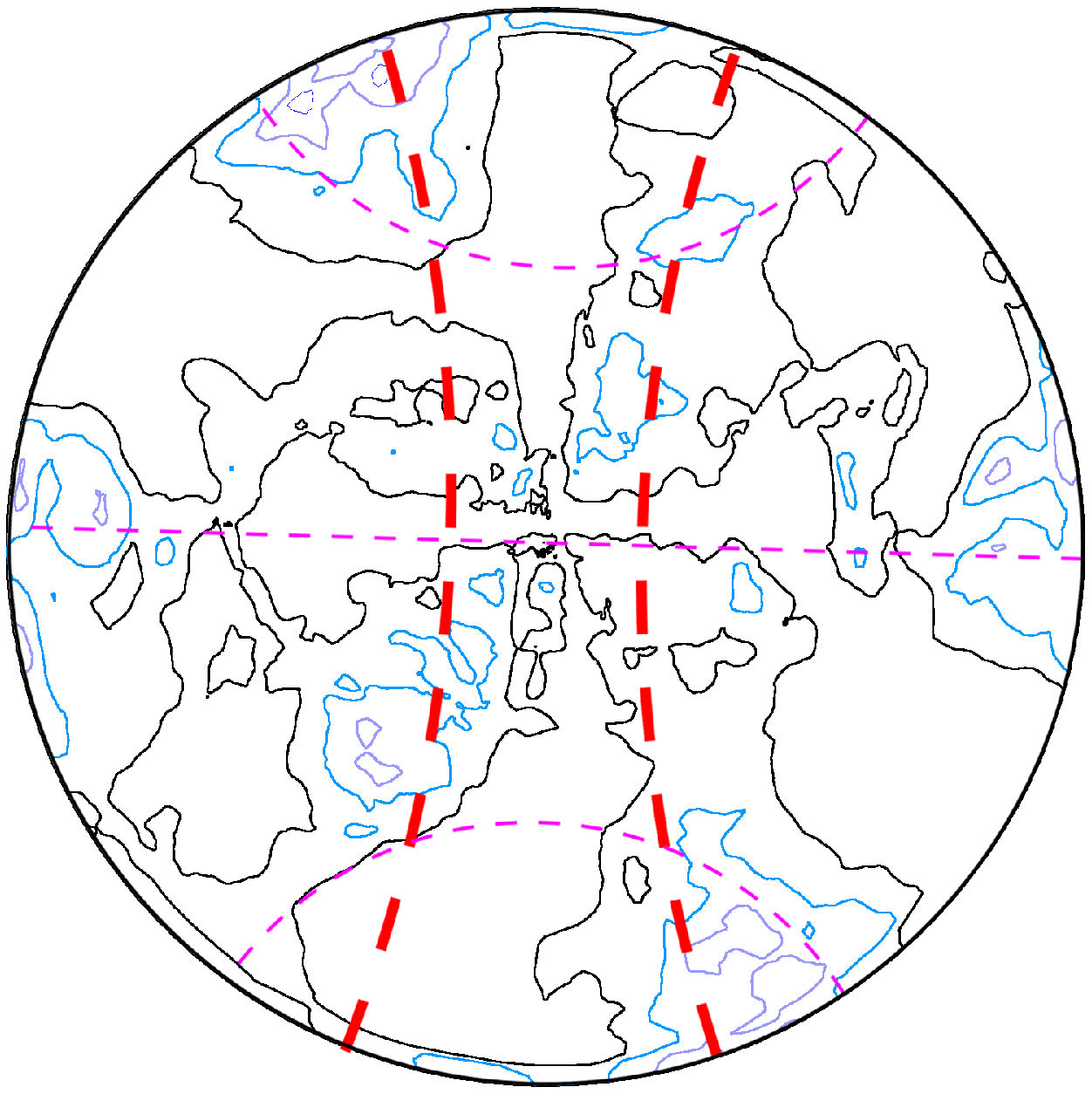}}
		\hfill\ 

		\hspace*{1ex}\hfill $\epsilon_2\sim1,500$\hfill\hfill  $\epsilon_2\sim400$\hfill\hfill  $\epsilon_2\sim1,000$ \hfill\hfill  $\epsilon_2\sim700$ \hfill\ 
		\end{minipage}
		\begin{minipage}[b]{0.07\textwidth}
		\includegraphics[width=0.9\textwidth]{TextureAxes}

		\includegraphics[width=0.9\textwidth]{PF_levels} 
		\end{minipage} 
		\caption{Pole figures showing the texture of the refractory element after second-step HPT deformation. Traces showing the ideal shear texture for BCC \cite{BaczynskiJonas1996} are overlaid on the pole figures for the 110 and 222 reflections of (c, g) Cu30Mo70 and (d, h) Cu20W80. No overlays are included for Cu57Cr43 and Cu50Mo50 since there were no regions of intensity $\ge$1.2. Long dashes (red in the colour version) indicate $\{hkl\}\langle111\rangle$ fibre textures ($D_1$,$D_2$,$E_1$ and $E_2$ fibres). Short dashed lines (magenta in the colour version) indicate $\{110\}\langle uvw\rangle$ fibre textures ($J_1$ and $J_2$).  The approximate shear strain imposed on each composite is indicated in the figure. 
\label{fig-pf-refractory-stage2}}
	\end{center}
\end{figure*}

Second step HPT deformation resulted in a weakening of the fibre texture of the refractory component, except in the case of Cu-W. Recalculated (110) and (222) pole figures for each of the refractory metal composites are presented in Figure~\ref{fig-pf-refractory-stage2}. For the composites which lacked a lamellar microstructure; Cu57Cr43 (Figs.~\ref{fig-APF_Cu57Cr43_110-stage2}--\ref{fig-APF_Cu57Cr43_222-stage2}) and Cu50Mo50 (Figs.~\ref{fig-APF_Cu50Mo50_110-stage2}--\ref{fig-APF_Cu50Mo50_222-stage2}) the texture was weak, with maximum intensities of 1.2, poorly defined and showed no clear correspondence with an ideal shear texture. (The texture strength data is summarised in Table~\ref{tab-texture-strength} in the supplementary material.)

The fibre texture was still evident in composites which retained a lamellar structure,  Cu30Mo70 (Figs.~\ref{fig-APF_Cu30Mo70_110-stage2}--\ref{fig-APF_Cu30Mo70_222-stage2}) and Cu20W80 (Figs.~\ref{fig-APF_Cu20W80_110-stage2}--\ref{fig-APF_Cu20W80_222-stage2}). The texture of the Cu30Mo70 composite became weaker for all reflections, with the maximum intensity in the (110) pole figure decreasing from 1.9 to 1.4. The rotation angle changed to $-14^\circ$, although this could be an artefact of the  broadening of the texture. Conversely the texture of the Cu20W80 composite became stronger, with the maximum intensity in the (110) PF increasing from 1.3 to 1.7. The tilt angle was decreased to 0 degrees although the fibres are not very well defined.

\section{Discussion}

Each of the refractory metal composites developed a lamellar structure during step 1 HPT deformation,  with the refractory layer spacing ranging from 400\,nm (Cu30Mo70) to 1.3\,$\mu$m (Cu57Cr43) at $\epsilon\sim$75 (See Table~\ref{tab-stereo}).  This microstructure is comprised of elongated refractory particles within the Cu matrix and bears a closer structural resemblance to the HPT deformed Cu-Ag alloys \cite{KormoutYang2015} than the more regular, planar composites produced by ARB \cite{BuetDubois2012,Beyerlein2013,CarpenterZheng2013}.

Deformation temperature has a strong influence on microstructural refinement. It has been shown that refinement of a Cu-25wt.\%W composite was noticeably more rapid during HPT deformation at room temperature than at 200--400$^\circ$C \cite{SabirovPippan2007}. Thus, the deformation temperature can have a strong influence on the particle refinement process.  In the present study, microstructure refinement was most extensive in the Cu57Cr43 alloy. The Cr has the highest DBTT of the three refractory elements (Cr, Mo, W) and thus it appear that the DBTT affects the ductility of the particles and hence their susceptibility for fragmentation by brittle fracture. For coarse grained refractory materials, the propensity for brittle fracture is more along the high angle grain boundaries. Thus, during the first-step of HPT deformation the coarser Cr particles break down at a faster rate. It has also been shown that introduction of dislocation sources by plastic deformation in tungsten can improve the ductility\cite{LinsmeierRieth2017}. The presence of elongated particles (Figure~\ref{fig-sem-cu-all}) indicates that the quasi hydrostatic state along with the applied shear strain can provide an appropriate stress state for ductilisation of refractory particles. However, this effect is most likely to be operating more effectively for Mo and W particles than for Cr due to the higher DBTT of the latter. Such a tendency was clearly observed during second-step HPT deformation.

In Cu30Mo70 and Cu20W80, containing 67 and 65 vol.\%  percentage of refractory particles, the lamellar structure with elongated refractory particles was maintained during second-step HPT deformation. However, for Cu57Cr43 and Cu50Mo50 composites, with refractory volume fractions of only 49\% and 46\% respectively, the elongated particle structure of the first-step HPT broke down into isolated and equiaxed hard particles embedded in the softer Cu matrix. This showed that apart from the DBTT of the particles the propensity for plastic deformation is also influenced by the volume fraction of the hard phase\cite{KormoutGhosh2017,KormoutPippan2016}. For a lower volume fraction of refractory particles the local deformation state acting on the particles possibly deviates significantly from the applied macroscopic deformation state leading to fragmentation of the lamellae.

The signature of such differences in the macroscopic and microscopic deformation state is evident from the crystallographic pole figures. The composites in which the lamellar structure was maintained developed a clear $\{hkl\}\langle111\rangle$ fibre texture (Figures~\ref{fig-pf-refractory-stage1} and \ref{fig-pf-refractory-stage2}). However, in the composites where the lamellar structure was decomposed, no prominent texture was observed. In this case, the equiaxed refractory particles are uniformly dispersed in the Cu matrix and rotate freely in response to the local deformation. This microstructure is also associated with the saturated hardness for Cu50Mo50, indicating that further refinement of the microstructure is significantly slowed down due to the unobstructed motion of hard refractory particles in Cu matrix. On the other hand, in Cu30Mo70 and Cu20W80, where a lamellar structure is maintained, further hardening was observed with increasing deformation (Figure~\ref{fig-hardness}). It is not clear till what extent of deformation the lamellar microstructure will be retained, as several equiaxed Mo and W particles started to form at higher strains (Figure~\ref{fig-stage-two-stem-Cu30Mo70}-\ref{fig-stage-two-stem-Cu20W80}). Although, at microscopic scale the lamellae still show signature of prominent texture, as observed by the SAD patterns (Figure~\ref{fig-stage-two-sad-Cu30Mo70}-\ref{fig-stage-two-sad-Cu20W80}), the bulk texture measured by synchrotron radiation shows weakening or spread about the ideal positions (Figure~\ref{fig-pf-refractory-stage2}). With a 120\,$\mu$m aperture and assuming a reasonable foil thickness of 100\,nm, the SAD pattern sampled a volume of 1100\,$\mu$m$^3$; five orders of magnitude less than the synchrotron X-ray beam. The variability in the shape and orientation of the refractory particles and their gradual fragmentation at higher strains, as observed in SEM (Figure~\ref{fig-sem-cu-all}) and STEM(Figure~\ref{fig-stage-two-stem}), leads to broadening and weakening of the texture due to averaging over the larger volume. A quantitative examination of the processes of plastic deformation and fragmentation of the refractory particles would require detailed stereological analysis which is beyond the scope of the present work although such a study is currently underway.

Although it was possible to produce nanostructured refractory metal composites via HPT, it should be noted that the refinement of the refractory particle size was substantially less than would be expected from the applied shear deformation.  
In an ideal case the thickness of the lamellae, $d$, is given by:
\begin{subequations}
\label{eqn-codeformation}
\begin{align}
d   =  \frac{d_0}{\gamma_1} 
 \mbox{~for 1 step HPT and} \\  
d = \frac{d_0}{~\gamma_1  \gamma_2} \mbox{for a 2 step process,}
\end{align}
\label{eqn-codeformation}
\end{subequations}
for shear strains of $\gamma_1$ and $\gamma_2$ where $\gamma_2$ is applied perpendicular to $\gamma_1$ and where the initial particle thickness is $d_0$\cite{BachmaierKerber2012}. The expected particle size and layer thicknesses are given in Table~\ref{tab-stereo}, together with the experimentally measured values. For completeness the refractory particle size is provided for both lamellar and isotropic composites, as the layer thickness or particle diameter, respectively.  After step 1 HPT deformation of Cu-Mo and Cu-W composites, the measured particle size is an order of magnitude larger than predicted, even though there was no indication of macroscopic slippage. The discrepancy is even greater during step 2 deformation where the strain imposed would theoretically reduce the layer thickness to (physically unrealistic) sub-nanometer scale, but where the measured values are actually two to three orders of magnitude greater.

\begin{table}
	\begin{center}
	\caption{A summary of the sample morphology and refractory metal particle size. ``AR" indicates the as-received condition. The measured refractory metal size is given as either the average Feret diameter for isotropic samples or the refractory layer thickness for lamellar samples.  For 2-step HPT deformation, $\gamma_1$ was calculated for the centre of the cylinder cut from the step 1 HPT disc. This was multiplied by the shear strain imposed during step 2 HPT. The expected particle size for ideal co-deformation is also given (See Equation~\ref{eqn-codeformation}).   \label{tab-stereo}}
			\begin{tabular}{p{4ex}lp{6ex}p{7ex}p{6ex}ll}
\toprule
 HPT Step & 
					&
Shear strain				&    
Morph\-ology						&
\multicolumn{2}{c}{Particle size}
&			\\

& 
&
$\gamma$ &
&
Meas. &
Ideal\\

& 
&
&
&
\multicolumn{2}{c}{(nm)} \\

\midrule
\multirow{4}{*}{AR}	  & 
Cu57Cr43 & 
\multirow{4}{*}{0} & 
\multirow{4}{*}{Isotropic}& 
4.1$\times10^4$ & 
\multirow{4}{*}{N/A}	
\\ 

 & 
Cu50Mo50 & 
 & 
&
5.0$\times10^3$ \\ 

& 
Cu30Mo70 & 
 & 
&	
3.8$\times10^3$ \\ 

 & Cu20W80 & 
 & 
&		
3.6$\times10^3$\\ 

\midrule
\multirow{4}{*}{1}	  & 
Cu57Cr43 & 
135 & 
\multirow{4}{*}{Lamellar} &				
1,300 &
300 &
\\ 

 & 
Cu50Mo50 & 
126 & 
&		
450& 
40 &
\\ 

& 
Cu30Mo70 & 
126 & 
& 
400 & 
30&		
\\ 

 & Cu20W80 & 
133 & 
& 
900 &
30 \\ 

\midrule

\multirow{4}{*}{2}	  & 
Cu57Cr43 & 
2600 & 
\multirow{2}{*}{Isotropic}& 		
14 & 
15 & 
\\ 

 & 
Cu50Mo50 & 
1.5$\times10^5$ & 
& 
17 &
0.03\\ 
\cmidrule{2-6}

& 
Cu30Mo70 & 
1.5$\times10^5$& 
\multirow{2}{*}{Lamellar}&
10$-$20& 
0.03			
\\ 

 & Cu20W80 & 
2.1$\times10^5$& 
&
20$-$50  & 
0.02
\\ 
\bottomrule
		\end{tabular}
	\end{center}
\end{table}

The ideal co-deformation of the two components assumed in Equation~\ref{eqn-codeformation} clearly does not occur in practice for Cu-Mo or Cu-W composites and it appears that strain tends to partition more to the softer Cu component. Co-deformation would result in a uniform reorientation and elongation of the refractory particles and  would also generate a strong texture. However, the SEM and STEM images (Figs.~\ref{fig-sem-cu-all} and ~\ref{fig-stage-two-stem}) show that the elongated particles are non-uniform in shape and have a spread of orientations, which becomes more pronounced after step 2 HPT. This is reflected in the development of texture in Cu30Mo70 and Cu20W80 during step 1 deformation (Fig.~\ref{fig-pf-refractory-stage1}) and its subsequent spread and weakening during step 2 HPT (Fig.~\ref{fig-pf-refractory-stage2}) as the lamellae become more convoluted and misoriented. The isotropic (step 2 HPT)  Cu50Mo50 composite also had a much larger Mo particle size than expected.

The microstructural refinement of the Cu-Cr composite was noticeably different to that of the Mo and W-based composites. The difference between ideal and measured layer spacing  was roughly a factor of 2 during step 1 deformation and the particle size after step 2 deformation was remarkably close that expected from the shear strain. As noted above, Cr has the highest DBTT of the refractory metals examined in this work, and this factor, combined with the larger initial particle size would allow brittle fracture to play a large role during the early stages of deformation. In addition, forced mixing can dissolve up to 32wt.\% Cu in Cr\cite{GuoRosalie2017a} and at higher strains this is likely to influence not only the volume fraction of the phases, but also their mechanical behaviour.  
Although it is possible to identify the DBTT and solubility as key parameters, the deformation behaviour of such lamellar composites is not well explained and this topic warrants further investigation.

Strong, well-defined orientation relationships between the co-deformed phases are a well-known characteristic of ARB Cu-Nb composites\cite{TothNeale1989,GhoshRenk2017} and influence the mechanical properties and radiation damage tolerance. However, the difficulty in obtaining the bulk texture data for the Cu phase makes it impossible to determine the orientation relationship between Cu and the refractory phases in present study and further local microscopic studies, such as the SAD data (Figure~\ref{fig-stage-two-sad}), are warranted. 

\section{Conclusions}
In summary, the present study showed that it is possible to circumvent the poor ductility of the group VI refractory elements (Cr, Mo, W) to produce ultrafine-grained composites in a Cu matrix by HPT deformation.  Cu and the refractory phases undergo limited co-deformation at room temperature and develop a discontinuous lamellar microstructure at an equivalent strain of $\sim$75. The refractory particles show considerable irregularity in their shape, size and orientation with respect to the shear plane. The refractory metals develop typical BCC $\{hkl\}\langle111\rangle$ fibre  textures, with a tilt to the tangential direction. The texture is stronger and more clearly defined in Cu-Mo and Cu-W than in Cu-Cr.

Further HPT deformation leads to the breakdown of the lamellar structure in Cu57Cr43 and Cu50Mo50, resulting in a saturated or near-saturated microstructure consisting of equiaxed, essentially randomly-oriented, nanometer-scale grains of the two phases. The grain sizes of Cr and Mo were measured as 14\,nm and 17\,nm, respectively. Cu30Mo70 and Cu20W80 (wt.\%) composites retained a nearly lamellar structure to extremely high applied strains and the refractory metal layer thickness was reduced to 10-20\,nm (Cu-Mo) or 20-50\,nm (Cu-W). Despite the microstructure being textured at a local scale, the irregularity of the refractory particle orientation did not allow for the development of strong, global texture.

The differences in the microstructure evolution of Cu57Cr43 and Cu50Mo50 from Cu30Mo70 and Cu20W80 composites can be attributed to the variation in DBTT and the volume fraction of the refractory metal. Thus, a wide range of microstructures can be generated by tuning the fundamental deformation behaviour and the volume fraction of the refractory elements.

\subsubsection*{Acknowledgements}
This work was conducted under FWF project 27034-N20 \textit{``Atomic resolution study of deformation-induced phenomena in nanocrystalline materials''}. PG gratefully acknowledges financial support from the European Research Council through project 340185 USMS. The starting materials were provided by Plansee, Austria. Synchrotron measurements were carried out at the Petra III beamline of the Deutsches Elecktonen-Synchrotron (DESY), Hamburg. The authors would like to thank Dr. Torben Fischer for assistance with the beamline experiments. The texture analysis was performed using code provided by Dr. Juraj Todt. SEM sample preparation was performed by S. Modritsch. The authors are grateful for valuable discussions with R. Pippan.

\newcommand{\noopsort}[1]{} \newcommand{\printfirst}[2]{#1}
  \newcommand{\singleletter}[1]{#1} \newcommand{\switchargs}[2]{#2#1}

\newpage
\renewcommand{\thesubsection}{S-\arabic{subsection}}
\renewcommand{\thefigure}{S-\arabic{figure}}
\setcounter{figure}{0}
\renewcommand{\thetable}{S-\arabic{table}}
\setcounter{table}{0}
\section*{Supplementary information}

\begin{table}[h!btp]
	\begin{center}
		\caption{Texture strength and tilt angle. The table shows the maximum texture intensity (MRD) for each system.  $\alpha$ is the amount by which the PF is rotated away from the ideal shear texture, measured from the (110) and (222) pole figures, respectively.. \label{tab-texture-strength} }
		\begin{tabular}{@{}lrrrr}
 &Cu-Cr	& 
\multicolumn{2}{c}{Cu-Mo}	& Cu-W	\\ \cmidrule{3-4}

hkl&
& 
Mo50 & 
Mo70 & 
\\ 
\midrule
  $\epsilon$: &~78 & 73 & 73 & 77 \\ \cmidrule(lr){2-5}

110		&
1.4	& 
1.8	& 
1.7	& 
1.3	 
\\

222		&
1.5	& 
2.1	& 
2.4	& 
1.5		\\ 

$\alpha(^\circ)$	&
11/19		&
12/11		&
8/8	&
10/14		\\

\midrule
  
$\epsilon_1$: &
$-$ &
50 & 
50 & 
100\\ 
$\epsilon_2$: &
1,500 &
1,000 & 
1,000 & 
700\\ 

\cmidrule(lr){2-5}

100		&
1.1	& 
1.2	& 
1.4	& 
1.8		\\ 

110		&
1.2	& 
1.2	& 
1.2	& 
1.7	 
	\\

111		&
1.2	& 
1.1	& 
1.3	& 
1.7		\\ 

$\alpha(^\circ)$	&
$-$ &
$-$ &
$-14$/$-14$		&
0/0		\\
\bottomrule
		\end{tabular}
	\end{center}
\end{table}

\begin{figure*}[h!btp]
	\begin{center}
		\includegraphics[width=0.3\textwidth]{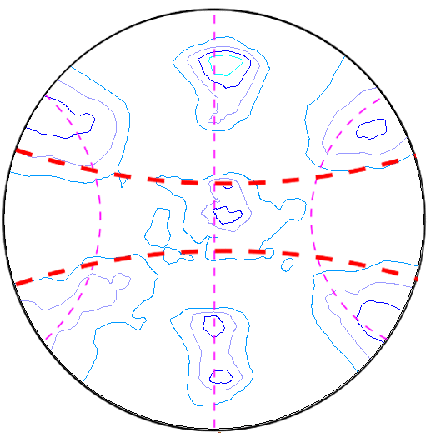}
		\includegraphics[width=0.063\textwidth]{TextureAxes} 		
		\includegraphics[width=0.12\textwidth]{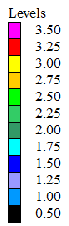} 

		\caption{111 pole figure for Cu in the Cu57Cr43 composite after step 1 HPT deformation. Maximum intensity was 2.3 MRD. 
Traces showing the ideal shear texture for FCC \cite{LiBeyerlein2005,TothNeale1989} are overlaid on the pole figures. Long dashes (red in the colour version) indicate $\{111\}\langle uvw \rangle$ fibre textures. Short dashed lines (magenta in the colour version) indicate $\{hkl\}\langle 110\rangle$ fibres. \label{fig-pf-cu57cu43_cu}}
	\end{center}
\end{figure*}

\end{document}